\documentclass[10pt,journal,compsoc]{IEEEtran}

\pdfoutput=1

\ifCLASSOPTIONcompsoc
  
  \usepackage[nocompress]{cite}
\else
  \usepackage{cite}
\fi

\usepackage{amssymb}
\usepackage{amsmath}
\usepackage[utf8]{inputenc}
\usepackage[T1]{fontenc}
\usepackage{graphicx}%
\usepackage{multirow}%
\usepackage{amsmath,amssymb,amsfonts}%
\usepackage{amsthm}%
\usepackage{mathrsfs}%

\usepackage[title]{appendix}%
\usepackage{xcolor}%
\usepackage{textcomp}%
\usepackage{manyfoot}%
\usepackage{booktabs}%
\usepackage{algorithm}%
\usepackage{algorithmicx}%
\usepackage{algpseudocode}%
\usepackage{listings}%
\usepackage{diagbox}
\usepackage{comment} 
\usepackage{tabularray}
\usepackage{subcaption}
\usepackage{nameref}
\usepackage{hyperref}
\usepackage{csquotes}
\usepackage{ulem}
\usepackage{array}
\usepackage{multirow}
\usepackage{ragged2e}
\usepackage{amsmath}
\usepackage{academicons}
\usepackage[utf8]{inputenc}
\usepackage[misc]{ifsym}
\usepackage{pifont}
\newcommand{\xmark}{\ding{55}}
\newcommand{\cmark}{\ding{51}}
 
\usepackage[utf8]{inputenc}
\usepackage{float}

\ifCLASSINFOpdf
\else
\fi

\begin{document}

\title{Flow-Based Detection and Identification of Zero-Day IoT Cameras}

\author{Priyanka Rushikesh Chaudhary and Rajib Ranjan Maiti}

\IEEEtitleabstractindextext{%
\begin{abstract}
  The majority of consumer Internet of Things (IoT) devices do not provide explicit mechanisms for network administrators to manage, monitor, and control them, which hinders the deployment of tailored IoT security policies. In particular, an administrator may notice, possibly by monitoring a range of IP addresses, that a new device has initiated its communication, but it is difficult to decide if it is an IoT device, in particular, a streaming IoT camera. 
In this paper, we design and develop a new system, called \textit{zCamInspector}, to identify a known set of IoT cameras using efficient supervised classifiers, called zCamClassifier, and to detect zero-day IoT cameras using efficient one-class classifiers, called zCamDetector. 
We have considered a total of about 40GB of network traffic categorized into three different broad types, Set I - consisting of six commercially available IoT cameras from different manufacturers - collected from our own IoT laboratory, Set II - consisting of five IoT cameras with about 1.5GB of open-sourced dataset, and Set III - consisting of four online audio/video conferencing applications and two video sharing applications - indicating non-IoT camera traffic. A set of 62 generic flow-based features are extracted with the CICFlowmeter tool.
Using the ExtraTree classifier, we show that top 10 features in these three datasets have a number of common features. 
zCamInspector uses seven supervised ML models, namely ET, DT, RF, KNN, XGB, LKSVM, and GNB, and four one class classifiers, namely OCSVM, SGDOCSVM, IF,and DeepSVDD. 
Our results show that zCamInspector achieves the highest accuracy of more than 99\% to identify the IoT cameras using XGB and false negatives as low as 0.3\% in both datasets. 
Compared to the state-of-the-art, zCamInspector using XGB achieves a better precision to identify some of the IoT cameras in both Set I and Set II.  
In detecting zero-day IoT cameras, our results show that accuracies are 93.20\%, 96.55\%, 78.65\% and 92.16\% using OneClassSVM, SGDOneClassSVM, IsolationForest,and DeepSVDD respectively, using trained samples of Set III (i.e., \texttt{Others} data set) and tested on test samples of Set I and Set II (i.e., \texttt{BITSHPC and UNSW} data sets).
We demonstrated the system's effectiveness in detecting zero-day IoT cameras under scenarios where all 11 devices are previously unseen, and when only one IoT camera is zero-day and rest are part of the network. 
When all the IoT cameras become zero-day, DeepSVDD (mean training accuracy as 96.03\% and mean testing accuracy as 74.51\%), showing dominating performance over remaining three models OneClassSVM (mean training accuracy as 85.64\% and mean testing accuracy as 60\% ), IsolationForest (mean training accuracy as about 89.75\% and mean testing accuracy as 62.79\%) and SGDOneClassSVM (mean training accuracy of about 43.21\% and mean testing accuracy as 67.05\%). 
Comparing with state-of-the-art, the detection of zero-day IoT cameras is shown for the first time in this paper and the results show good performance of zCamInspector by achieving more than 95\% accuracy in case of certain cameras, like Spy Clock camera.

\end{abstract}

}

\maketitle

\IEEEdisplaynontitleabstractindextext

\IEEEpeerreviewmaketitle

\IEEEraisesectionheading{\section{Introduction}\label{sec:introduction}}

Several recent market studies have revealed that the usage and prices of indoor and/or outdoor IoT surveillance cameras will occupy a huge volume that is worth 13.2 billion USD by 2032 \cite{market1Cam_2024,market11Cam_2023}.
According to NIST \cite{NISTIR8288, NISTIR8228}, a vast majority of IoT devices cannot be accessed, managed, and monitored in the same way as conventional IoT devices, like Desktop, Laptop, Printers, Smartphones, and Tablets. 
This is because the IoT devices are often considered as \enquote{black boxes} to the network administrators, having no information about their states, meaningful identities, or the external systems they interact with. 
This is underpinned by the fact that rule-based attack detection mechanisms \cite{visoottiviseth2020IEEE, chaabouni2019IEEE, Snort, Suricata} are challenged due to the adoption of a same set of cloud service providers by a larger number of IoT vendors \cite{IP_awscloud}.
Among others, IoT cameras pose direct threats to user privacy where face images, behaviors, gestures, personally identifiable information (PII), phone numbers, home addresses, or home assets can be directly exposed to attackers \cite{alharbi2018IET}. 
Certain case studies on specific IoT cameras (like VibeX Onvif YY HD IoT camera \cite{Abdalla2020ISDFS} and Belkin NetCam camera \cite{tekeoglu2015IEEEICCCN}) and IoT Hub (like SmartThingsHub \cite{fernandes2016IEEE} that can integrate IoT cameras) reveal that the risk of security breach is higher in IoT cameras compared to other IoT devices. 
Due to a lack of standardized mechanisms, the network administrators face humongous challenges to manage these devices using a centralized management system and this is aggravated by the scale of IoT devices in the network.  
In this paper, we address the problem of managing such IoT cameras by analyzing and identifying their network flows without relying on any IP address or transport ports or breaking encryption.

\begin{figure}[h]
\centering
\includegraphics[width=\linewidth]{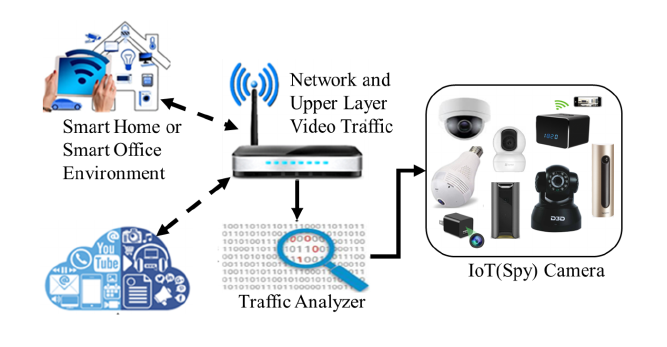}
  \caption{A motivating example.}
  \label{fig:motivate}
\end{figure}
We motivate our study by using a smart home scenario (Fig. \ref{fig:motivate}), similar to that as envisioned in \cite{Arunan2018IEEE, PratibhaIET2021}. The smart home has a number of traditional computing devices, like a desktop or a smartphone, running certain online audio/video conferencing applications like Meet and video sharing applications like YouTube. Note that these applications do generate video flows, and these flows are acceptable usually.
Also, certain IoT devices like smart bulbs and smart doorbells may be used in the environment. 
However, installation of a multi-functional IoT device, like smart bulb cum spy camera (Spy Bulb) \cite{lightbulbcumcamera}, is not expected if not authorized apriori. 
In this scenario, we put forward three broad  questions related to cyber security and management of IoT device in general and IoT cameras in particular. First, is it possible to identify the network flows of a particular IoT (spy) camera in an IoT network for better inventory management? 
Second, is it possible to detect an IoT camera whose flow patterns are not known apriori, i.e., zero-day IoT camera, in the presence of network flows arising from online audio/video conferencing, like Meet, Teams, and Skype, or video sharing applications, like YouTube and Prime, leading to imposing organization specific cyber security policies?  
Third, assuming the IoT devices can be behind NATs, is it possible to make the identification of IoT devices IP addresses and transport ports agnostic of both sources and destinations, i.e., agnostic of external systems that the devices interact with \cite{Mazhar2020arXiv}. 
Each of these questions is related to some of the central challenges of IoT management and cyber security as per NIST \cite{NISTIR8228}.
To address these questions, we have overcome three main challenges. 
First, we have considered a large volume of network traces of 11 IoT cameras collected from our own laboratory setup and freely available real-world datasets, 4 online audio/video conferencing applications, and 2 online video-sharing applications, having overlapping cloud service providers. 
Second. we have carefully prepared a large volume of datasets on a case-by-case basis to consider zero-day network traffic of IoT cameras and assumed a specialized machine learning model to detect zero-day IoT cameras.  
Third, we have segregated the network traffic into smaller flows and extracted a large set of flow-based generic features without utilizing IP addresses and transport ports. 

In this paper, we propose a new system called \textit{zCamInspector} that substantially surpasses the capabilities of recently proposed  \texttt{iCamInspector}  \cite{iCamInspector2024BuildSec} \footnote{We make portion of code base along with sample network traces available online \cite{iCamInspector_githubRepo}}. 
In brief, iCamInspector works in two stages, Stage-I and Stage-II. 
Stage-I runs \texttt{vfClassifier} that classifies any network flow into one of four categories, namely, \texttt{ConfFlows} - denoting that the flows belong to online audio/video conferencing applications, e.g., Skype, Zoom, Meet and Teams, \texttt{ShareFlows} - denoting that the flows belong to video sharing applications, e.g., YouTube and Prime, \texttt{IoTCamFlows} - denoting that the flows belong to the category of IoT camera, e.g., D3D, Netatmo, and Spy Clock, and \texttt{Others} - denoting that the flows belong to any other application that does not produce audio/video data. 
Stage-II runs any of three additional supervised classifier modules, namely, \texttt{ConfDetector}, \texttt{ShareDetector} and \texttt{IoTCamDetector} - identifying that the flow belongs to a specific conferencing application, share application, and IoT camera respectively. 
All these classifiers in iCamInspector have used only classification and regression tree (CART) model. 
We consider the results in \cite{iCamInspector2024BuildSec}  as a baseline in this paper.

The capabilities of zCamInspector  are supported by three factors. First, an open-sourced data set is considered together with our own dataset that demonstrates the efficacy of our proposed system outside our laboratory setup.  
Second, four one-class classifiers are considered to enhance and contrast the performance of detecting zero-day IoT cameras, thus calling our system as zCamInspector.   
Third, zCamInspector considers an extended set of supervised ML models, namely, Extra Trees (ET), decision tree (DT), random forest (RF), k-nearest neighbor (kNN), extreme gradient boosting (XGB), Guassian Naive Bayes (GNB) and linear kernel SVM (LKSVM) that identifies a particular IoT (spy) camera and contrasts the performance with state-of-the-art including \cite{iCamInspector2024BuildSec}, and four one-class classifiers, namely OneClassSVM, SGDOneClassSVM, IsolationForest and DeepSVDD for detecting zero-day IoT cameras in three cases; all IoT cameras are zero-day, all but one is zero-day and only one is zero-day.

For the work in this paper, we have considered a total of 40GB of network traces (basically pcap files) that includes IoT camera traffic of about 1.49GB open-sourced pcap files and 14.96GB of pcap files from our own laboratory setup.  
Similar to iCamInspector \cite{iCamInspector2024BuildSec}, zCamInspector does not attempt to decrypt any network packet nor use IP addresses or port numbers as features. 
Overall, the dataset is divided into three categories. Set I \textit{(BITSPHC)} -- contains the IoT camera traffic from our laboratory setup, and the vast majority of packets contain video data in their payloads. 
Set II \textit{(UNSW)} -- contains the IoT camera traffic from an open-sourced data set, where the vast majority of packets carry only handshake packets. 
Set III \textit{(Others)} -- contains a mix of pcap files representing other applications, like online audio/video applications, e.g., Skype, Teams, Zoom, and Meet, video sharing applications, e.g., YouTube and Prime, and other applications that do not produce audio/video data. 
We have used \texttt{CICFlowmeter} \cite{cicflowmeter}, an open-sourced tool, to extract 62 generic flow-based features. We have investigated the performance of the ML models using all of 62 features and using only top 10 important features that are obtained using Extra Tree classifier.
We have performed an extended set of experiments to detect zero-day IoT cameras and contrasted the performance of zCamInspector with state-of-the-art in both identifying the IoT cameras and detecting the zero-day IoT cameras. 
We summarise our contributions as:
\begin{enumerate}
    \item We have designed and developed a system called zCamInspector to efficiently identify IoT cameras by using advanced supervised machine learning models and to detect zero-day IoT cameras by using one-class classifiers, all using network flow-based features to make the system agnostic of any IP address range or transport ports. 
    \item Using the Extra Tree Classifier, we show that only the top ten important features can achieve a better model performance compared to baseline results in \cite{iCamInspector2024BuildSec} to identify the IoT cameras. Further, the sets of top ten features in the three datasets have significant variations, and hence, we consider our dataset \footnote{A sample of data set is already made available for the community in \cite{zCamInspector_githubRepo}, the complete dataset will be released upon the acceptance of this journal version.} to be a new contribution for the research community for further investigations.  
    \item Our analysis shows that the XGB classifier achieves the highest accuracy in Set I and the combined dataset of Set I and Set II, which is about 4\% higher than the CART model.
    Further, we show that the performance of DeepSVDD defeats other three models OneClassSVM, IsolationForest and SGDOneClassSVM in detecting any of the zero-day IoT cameras. In particular, we show that DeepSVDD achieves a mean accuracy of about 96.03\% and testing accuracy as 74.51\%. 
    
    \item Finally, we show that zCamInspector 
    achieves a precision of over 97\% using XGB, which is significantly higher than the state-of-the-art works in identifying the IoT cameras in supervised classification. 
\end{enumerate}

In the rest of the paper, 
Section \hyperref[sec:relwork]{"Related Work"} describes the works in classifying IoT cameras. 
Section \hyperref[sec:sysem-architecture]{"Sysetm Design and Implementation"} provides the design and implementation of iCamInspector.
Section \hyperref[sec:dataset]{"Experimental Setup and Dataset"} summarizes the experimental setup with datasets used in this paper.
Section \hyperref[sec:detection-iot-camera]{"Detecting IoT Camera Flows"} classifies IoT camera flows, detection of zero-day IoT cameras with a runtime performance followed by the conclusion in Section \hyperref[sec:conclusion]{"Conclusion"}.

\section{Related Work}
\label{sec:relwork}
\subsection{Classifying IoT Devices}

\begin{table*}[hbtp]
\centering
\caption{Comparison with Existing Work. A customized dataset indicates a dataset created in one's own labs.}
\label{tab:rel_table}
\begin{tabular}{ccccccc}
\hline
\multirow{2}{*}{Research Paper} & \multirow{2}{*}{Dataset} & \multirow{2}{*}{\begin{tabular}[c]{@{}c@{}}Flow\\ based\\ Features\end{tabular}} & \multicolumn{2}{c}{Classification} & \multirow{2}{*}{\begin{tabular}[c]{@{}c@{}}Detection of \\ Spy Camera\end{tabular}} & \multirow{2}{*}{Techniques} \\ 
\cmidrule(l){4-5}
 &  &  & \begin{tabular}[c]{@{}c@{}}IoT \\ Devices\end{tabular} & \begin{tabular}[c]{@{}c@{}}IoT \\ Cameras\end{tabular} &  &  \\ \hline \hline
A. Sivanathan et. al.\cite{Arunan2018IEEE} & UNSW & \xmark & \cmark & \xmark & \xmark & \begin{tabular}[c]{@{}c@{}}Machine learning using\\ Naive Bays and Random\\ Forest classifier\end{tabular} \\ \hline
P. Khandait et. al.\cite{PratibhaIET2021} & UNSW & \xmark & \cmark & \xmark & \xmark & \begin{tabular}[c]{@{}c@{}}Deep Packet Inspection\\ based Network Flow \\ Classifier\end{tabular} \\ \hline
J. Hao et.al. \cite{heoACSAC2022} & \begin{tabular}[c]{@{}c@{}}Customized\\ Dataset\end{tabular} & \xmark & \xmark & \cmark & \cmark & \begin{tabular}[c]{@{}c@{}}Encrypted traffic\\ Analysis\end{tabular} \\ \hline
Y. Cheng et.al. \cite{Cheng2018ASIACCS} & \begin{tabular}[c]{@{}c@{}}Customized\\ Dataset\end{tabular} & \xmark & \xmark & \cmark & \cmark & \begin{tabular}[c]{@{}c@{}}Analyzing traffic\\ features, Android\\ platform\end{tabular} \\ \hline
\begin{tabular}[c]{@{}c@{}}zCamInspector\\ (Our Approach)\end{tabular} & BITSPHC & \cmark & \xmark & \cmark & \cmark & \begin{tabular}[c]{@{}c@{}}Machine Learning\\ Extra Tree Classifier\\ OneClassSVM, \\SGDOneClassSVM \\ Isolation Forest\\ DeepSVDD\end{tabular} \\ \hline
\end{tabular}
\end{table*}

A number of existing studies have investigated the feasibility of classifying and detecting IoT devices in general and streaming IoT devices, i.e., IoT cameras, in particular, possibly because IoT cameras pose direct security and privacy threats. 
Both the users and the developers communities of IoT devices in general, and IoT cameras in particular, are less careful about the security and privacy features of the devices \cite{alharbi2018IET}, and this attracts a significant attention from research community across the globe. 
We categorize the existing relevant works into two broad classes: classifying IoT cameras and detecting zero-day threats in IoT devices.

In the first category, the work in \cite{heoACSAC2022} proposes a system, called \texttt{Spy Camera Finder (SCamF)}, to classify IoT camera traffic, then to detect the presence of a streaming IoT spy camera by using Wi-Fi traffic, i.e., MAC layer traffic, and localize such IoT spy camera. 
SCamF classifies camera
and non-camera traffic by using flow level features
such as traffic volume, inter-packet time interval, fragmentation unit (FU) rate, and frame per second (FPS). 
Similarly, IoT-Praetor framework \cite{wang2020iot} can be used to  detect malicious behavior of IoT devices using new device usage description (DUD) model to construct the behavior specification of IoT devices, including the desired communication behaviors and interaction behaviors.
The DecoyPort framework  \cite{Kim2007Springer} uses a classical technique of interception and port redirection to steer attackers to honeypots so that the attacker behavior can be predicted by analysing the logged data and network traffic, in particular using Source IP address, destination port, payload contents, protocols etc.
Similarly, Siphon framework \cite{guarnizo2017siphon} can be used to derive attackers intent by building a scalable, high-fidelity honeypot infrastructure using traffic forwarding via reverse SSH tunnels that mimics real physical IoT devices, like IoT cameras, printer etc., and collect data rich in behavioral aspects of the attacker. 

Further, the detection of hidden IoT camera using wireless traffic (i.e., MAC layer traffic) has been proposed in \cite{cowan2020detecting, Cheng2018ASIACCS}, and it is relatively more challenging due to the lack of access of IP addresses and transport ports, if Wi-Fi is secured.  
Classification of IoT devices via the decryption of encrypted Wi-Fi traffic that exploits the vulnerability in the security settings of IoT devices was proposed in \cite{Zhang2018ACM}.
However, as the network traffic (i.e., Network and Transport layer traffic) can be accessed directly from the networking infrastructure, the use of features from network traffic can be widely observed in the existing works \cite{detectingIoT_IEEE2020, SivanathanA2018IEEE, Gordon2021IEEE}.  Such works have been extended to industrial IoT systems as well \cite{NawrockiM2020IEEE}, and the results show that such extensions are not trivial. 
Further, the use of application layer information to classify IoT devices has also been explored with the help of natural language processing (NLP) on application data \cite{thomsen2020smartlampsmartcam}.  

\subsection{Detecting Zero-Day Threats}
Among the works in the second category, the work in \cite{sharma2018arXiv} introduces a context graph–based, distributed diagnosis framework to detect and mitigate zero‑day threats by utilizing distributed trust verification between central service providers and local IoT devices. When anomalies are detected, a critical data-sharing protocol restores trust across nodes which improves efficiency over centralised systems. 
For example, DIoT framework \cite{nguyen2019IEEE} uses federated self-learning based anomaly detection mechanism for emerging unknown threats. It uses device‑type profiling via behavioral models and federated learning. Such a framework can be more efficient to detect unknown malware, like Mirai, and be generalizable to detect zero‑day alike behavior. 
Similarly, ZDT method \cite{redino2022IEEE} uses dual-autoencoder models on network flows and graph features to detect novel threats which results in low‑false‑positive zero‑day detection in enterprise networks. Recently, the work in \cite{Agbedanu2025MDPI} has proposed to detect zero-day using adaptive k‑NN with federated learning and ensemble techniques. 
Further, the researchers have proposed a hybrid detection approach for zero-day threats that integrates machine learning (ML) and deep learning (DL) techniques \cite{asaduzzaman2022IEEE}, \cite{he2023image}, \cite{armijos2023IEEE}. 
Notably, this framework achieves near 99\% detection rate while maintaining low false positive and false negative rates \cite{he2023image}. 
The \cite{saurabh2025Springer} introduced a hybrid intrusion detection system (IDS) that leverages both supervised and unsupervised ML methods to identify known and novel threats. 
The research shows the effectiveness of ensemble methods to detect zero-day threats in IoT networks \cite{zhang2023IEEE}, \cite{guo2023IEEE}, \cite{nkongolo2023Springer}. By integrating multiple base anomaly detectors built on conventional machine learning algorithms, ensemble methods can achieve high detection accuracy even in the absence of labeled attack data. Notably, ensemble techniques utilizing Random Forest (RF) and Extreme Gradient Boosting (XGB) have emerged as top-performing approaches for zero-day attack detection in IoT environments. These models not only outperform traditional methods but also significantly enhance the overall performance of machine learning-based detection systems \cite{nkongolo2023Springer}. An ensemble deep learning methods also used in detecting zero-day threats along with intrusion detection systems (IDS). The proposed IDS \cite{ahmad2022Elsevier} was trained using four benchmark datasets and evaluated against previously unseen attacks to assess its effectiveness in detecting zero-day threats.

\subsection{Positioning The Work in This Paper}
Our work in this paper is positioned in the intersection of these two types existing works, where our first aim is to detect any zero-day IoT camera and then classify the known IoT devices based on the flow-based features of their network traffic. 
In particular, we consider the features that do not depend on IP address and transport number so that the classification can be performed even if the IoT devices behind any layer of NATed routers.  
Such classification of IoT devices can be very handy in scenarios where certain access control mechanisms are employed at a central location, possibly at a gateway.  
Both the tasks are achieves by using a same set of features so that the overall framework can be optimized and a uniformity can be maintained. 
In this paper, we introduce a new IoT dataset that not only contains the header information, but also the payload, and we have extensively address the problem of detecting zero-day streaming IoT camera that is under-studied in the existing literature.
The summary of comparison of the relevant existing works is shown in Table \ref{tab:rel_table}.

\section{System Design and Implementation}
\label{sec:sysem-architecture}

\subsection{The Design of the System Model} 
\begin{figure*}[h!]
\centering
\includegraphics[width=\linewidth]{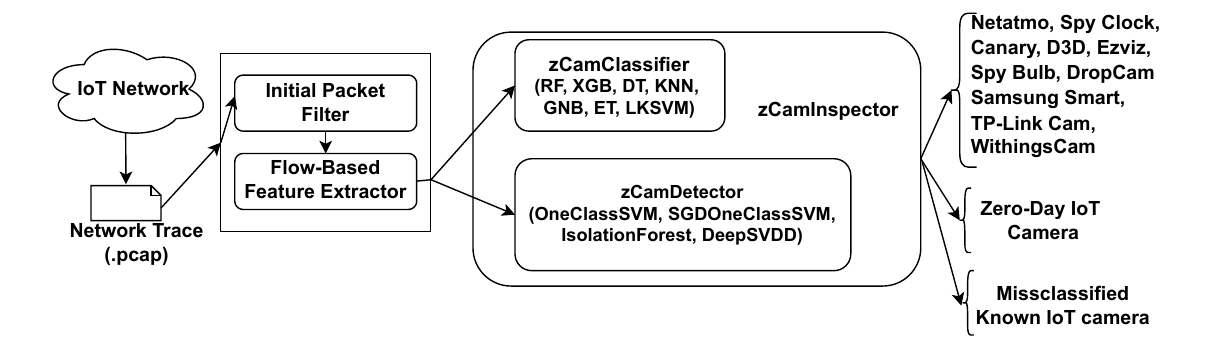}
\caption{Architecture of our proposed system zCamInspector.}
\label{fig:architecture}
\end{figure*}

Fig. \ref{fig:architecture} shows a high-level design of the proposed system, called \texttt{zCamInspector}, that is applicable to detect IoT cameras in a smart environment. 
zCamInspector accepts as input the network and upper layer traces (i.e., \texttt{.pcap} files) collected at a router or a switch having a port-mirroring option.
This input, after an initial filtering, is passed through a flow-based feature extractor to extract a set of flow-based features. 
Note that none of the features contains an IP address or transport ports, thus making the system NAT (Network Address Translation) agnostic.  
zCamInspector mainly consists of two classifiers, namely \texttt{zCamClassifier} and \texttt{zCamDetector}. 
The features are fed to a classifier, called \texttt{zCamClassifier}, and another, zero-day IoT camera detector called \texttt{zCamDetector}.
zCamDetector basically tries to detect any IoT camera that has not been seen before, and \texttt{zCamClassifier} detects a particular known IoT camera based on its past patterns. 
zCamDetector uses OneClassSVM, SGDOneClassSVM, Isolation Forest and DeepSVDD to find zero-day IoT cameras (outlier detection). 
zCamClassifier uses RF, XGB, DT, KNN, GNB, ET and LKSVM models to detect known IoT camera.
This system can be used by a network administrator to constantly and automatically monitor the networking environment where an attacker may activate an IoT camera at any time in the environment without authorization, causing a privacy and/or security threat.

\subsection{Threat Model}
An attacker can pose privacy and security threats by either implanting any number of IoT cameras or arbitrarily activating any IP camera service in existing IoT devices. The devices that offer multiple services include but are not limited to, smart plug-cum-camera and smart bulb-cum-camera. 
It is assumed that any of these devices avail Internet connectivity via existing routers, either Wi-Fi or Ethernet, and the router is under the control of the administrator.
The attacker does not have the capability to compromise such a router. Further, the attacker does not have the capability to compromise the defense mechanism adopted by the administrator, like the firewall or an ML-based anomaly detector that monitors and controls the ingress or egress network traffic.
However, the attacker can impersonate an existing IoT device by using its IP or MAC address and can make use of those cloud servers whose IP addresses are already white-listed by the administrator. 

Once an IoT camera is installed or activated, an attacker can exfiltrate the video stream at any arbitrary time, directly invading the privacy of target users. 
In the case of existing devices, the attacker can activate any number of IoT cameras simultaneously, fix the focus to an arbitrary area, zoom in/out for a detailed capture, and so on, without the knowledge of the owner. 
Though any change in the parameters of the camera by an attacker can be an attack, we consider that an IoT camera itself is a threat as soon as it starts streaming.

\subsection{Defender Model}
A network administrator employs a number of defense mechanisms, including firewalls at the router and additional computing devices like a gateway to defend its network under control (similar to that in \cite{Sun2019AutomatedID, IbbadTNSM2020}). 
At the same time, we envision that a defender can employ a mechanism for  monitoring and controlling the network and above-layer traffic. 
Given the diversity and dynamics of the networked applications and the computing devices (including but not limited to traditional desktops, laptops, smartphones, and IoT devices, including IoT cameras), the defender is interested in detecting known or unknown IoT camera streams irrespective of whether the traffic is encrypted or not. 
To address such a challenging task, the defender can make use of zCamInspector, which can raise flags to indicate either existing IoT camera is streaming or a zero-day IoT camera is implanted. 
The defender aims to minimize the false negative alarms, i.e.,  an actual zero-day IoT camera stream should not go undetected, or a known IoT camera should not be misclassified. 
In case of false positive alarms, the administrator can simply inspect further to rule out the streams on a case-by-case basis.   

\subsection{Implementation of zCamInspector}
\texttt{zCamInspector} has been implemented using freely available well-known tools that are compatible with Python3. 
In particular, feature extraction from the new dataset is performed using \texttt{CICFlowmeter}\cite{cicflowmeter}. 
It is developed at the Canadian Institute for Cybersecurity (CIC) research lab and University of New Brunswick (UNB), Canada. We use this tool, to extract 77 flow-based features, \texttt{SciKit Learn} package for machine learning algorithms and \texttt{Pandas} for data preparation and preprocessing. 

\subsubsection{Flow-based Feature Extraction}
CICFlowmeter identifies a flow by a 6-tuple of [FlowID, SourceIP, DestinationIP, SourcePort, DestinationPort, Protocol], where FlowID is a unique integer, SourceIP (SourcePort) and DestinationIP (DestinationPort) are IPv4 address (transport port number) of the source and destination hosts, and Protocol is the value of \textit{protocol} field in IP header. 
Usually, the flow is directional.
A flow in TCP starts and ends with SYN and FIN packets, respectively, and in UDP, it is based on a time duration. However, CICFlowmeter limits a flow to a default time window of 600s in both TCP and UDP.
The tool extracts 77 flow-based features from each such flow and stores them in a \texttt{csv} file; one record has 77 (features) + 6 (6-tuple) + 1 (label) = 84 values, where the \textit{label} is predefined.

\subsubsection{Feature Pre-processing and Selection}
The features extracted by CICFlowmeter are generic in nature, and it is not necessary for every type of application to generate a valid set of values for each of these features. 
Hence, at first, we apply a set of standard techniques, like standard deviation, to discard any feature that has a fixed values. 
Then, to obtain a set of important features specific to the application in this paper, we apply an Extra Tree-based classifier, considering it is robust to any kind of noise in the dataset. 

\subsubsection{Machine Learning Model Selection}
To select an effective model from a range of models, we consider a number of one class classifiers, namely one-class support vector machine (OCSVM), SGD-one-class support vector machine (SGDOCSVM), Isolation Forests (IF) and Deep Support Vector Data Description (DeepSVDD) for the detection of zero-day IoT cameras.
For IoT camera detection, we consider a number of supervised machine learning models, including Decision tree (DT) model, Extra Tree (ET), Gaussian naive Bayes (GNB), random forest (RF), k-nearest neighbor (kNN), extreme gradient boosting tree (XGB) and linear kernel support vector machine (LKSVM). 

\begin{itemize}
    \item \textbf{One-Class Support Vector Machine (OCSVM) :} OC-SVM is a classical unsupervised anomaly detection technique that aims to learn a decision boundary that encloses the majority of the data points (assumed to be normal) while isolating potential anomalies. It maps the input data into a high-dimensional feature space using a kernel function and finds the hyperplane that separates the mapped data from the origin with maximal margin. The optimization problem is formulated as:
    \begin{equation}
\min_{w, \rho, \xi} \frac{1}{2} \|w\|^2 + \frac{1}{\nu n} \sum_{i=1}^n \xi_i - \rho
\end{equation}
\begin{equation}
\text{subject to: } (w \cdot \phi(x_i)) \geq \rho - \xi_i, \quad \xi_i \geq 0
\end{equation}

where $\phi(x_i)$ is the feature mapping, $w$ is the weight vector, $\rho$ is the offset, $\xi_i$ are slack variables, $\nu \in (0,1]$ is a parameter controlling the trade-off between the volume of the hypersphere and the number of outliers, and $n$ is the number of training samples \cite{scholkopf2001NeuralCom}.

\item \textbf{Stochastic Gradient Descent One-Class SVM (SGDOCSVM) :} SGDOCSVM is a scalable variant of OC-SVM that uses stochastic gradient descent to optimize the one-class objective over large datasets. It replaces the conventional quadratic programming solver with iterative updates, making it suitable for online learning and high-dimensional data. The objective is reformulated using hinge loss:

\begin{equation}
\min_{w} \frac{1}{2} \|w\|^2 + \frac{1}{\nu n} \sum_{i=1}^n \max(0, \rho - w^\top x_i)
\end{equation}

The model updates the weight vector $w$ using mini-batch gradients computed only on misclassified or marginal points. This efficient optimization enables the deployment of OC-SVM on streaming and large-scale anomaly detection tasks \cite{gornitz2013ACMSIGKDD}. 
 
\item \textbf{Isolation Forest (IF) :} Isolation Forest (IF) is an ensemble-based method designed specifically for anomaly detection. Unlike density-based or distance-based methods, IF focuses on isolating anomalies by recursively partitioning the data using randomly selected features and split values. The intuition is that anomalies are few and different, and thus are easier to isolate in fewer splits. The anomaly score of a point $x$ is defined as:

\begin{equation}
s(x, n) = 2^{-\frac{E(h(x))}{c(n)}}
\end{equation}

where $h(x)$ is the path length of $x$ in the tree, $E(h(x))$ is its expected value across all trees, $n$ is the sample size, and $c(n)$ is the average path length in a binary search tree used for normalization. Shorter path lengths correspond to higher anomaly scores \cite{liu2008IEEE}.

\item \textbf{Deep Support Vector Data Description (DeepSVDD)}
DeepSVDD extends the Support Vector Data Description (SVDD) approach by integrating deep neural networks to learn representations that enclose normal data within a minimal-volume hypersphere in a latent space. The network is trained to map inputs to a representation close to a predefined center $c$ in the output space. The objective function minimized is:

\begin{equation}
\mathcal{L}(\theta) = \frac{1}{n} \sum_{i=1}^{n} \|f_\theta(x_i) - c\|^2
\end{equation}

where $f_\theta(x_i)$ denotes the neural network output for input $x_i$, parameterized by weights $\theta$, and $c$ is the center of the hypersphere. During inference, data points with distances greater than a chosen threshold are flagged as anomalies. A soft-boundary version of DeepSVDD incorporates a margin controlled by a parameter $\nu$ to allow a fraction of the training samples to lie outside the hypersphere \cite{ruff2018dsvdd}.
\end{itemize}


Because a defender would like to minimize the false rates, our experiments would lead to discovering a model that fits better to the problem being addressed in this paper and the datasets for the proof of concept. 
Unless specified, we use the sci-kit learn package along with the default settings of the hyperparameters in each of these models.
Finally, in order to detect zero-day IoT cameras, we use the OCSVM, SGDOCSVM, IF and DeepSVDD with a single class of data for training, and the test data consists of a mix of samples from known class and samples from the unknown class.


\section{Experimental Setup and Data Set}
\label{sec:dataset}

\begin{table*}[h]
\centering
\caption{Summary of network traffic and flows.}
\label{tab:camera-data-set}
\begin{tabular}{clccccc}
\hline
Set & \begin{tabular}[c]{@{}l@{}}Collection\\ Site\end{tabular} & IoT Camera (Model No.) & \begin{tabular}[c]{@{}c@{}}Size\\ (GB)\end{tabular} & Packet Count & \begin{tabular}[c]{@{}c@{}}\#Total\\ Flows\end{tabular} & \begin{tabular}[c]{@{}c@{}}Post-preprossing\\ \#Flows\end{tabular} \\ \hline \hline
\multirow{6}{*}{\begin{tabular}[c]{@{}c@{}}Set I\\ (BITSPHC)\end{tabular}} & \multirow{6}{*}{\begin{tabular}[c]{@{}l@{}}Own \\ Laboratory\end{tabular}} & Netatmo (NSC01-US) & \multirow{6}{*}{14.96} & \multirow{6}{*}{25,675,111} & 943 & \multirow{11}{*}{119932} \\ \cline{3-3} \cline{6-6}
 &  & Spy Clock (V380) &  &  & 395 &  \\ \cline{3-3} \cline{6-6}
 &  & Canary (CAN100USBK) &  &  & 421 &  \\ \cline{3-3} \cline{6-6}
 &  & D3D (D8801) &  &  & 1733 &  \\ \cline{3-3} \cline{6-6}
 &  & Ezviz (CS-CV246) &  &  & 2586 &  \\ \cline{3-3} \cline{6-6}
 &  & Spy Bulb (H18E) &  &  & 3922 &  \\ \cline{1-6}
\multirow{5}{*}{\begin{tabular}[c]{@{}c@{}}Set II\\ (UNSW)\end{tabular}} & \multirow{5}{*}{\begin{tabular}[c]{@{}l@{}}Open\\ Source\end{tabular}} & Dropcam & \multirow{5}{*}{1.49} & \multirow{5}{*}{6,586,379} & 6577 &  \\ \cline{3-3} \cline{6-6}
 &  & Netatmo &  &  & 17197 &  \\ \cline{3-3} \cline{6-6}
 &  & Samsung SmartCam &  &  & 61543 &  \\ \cline{3-3} \cline{6-6}
 &  & TP-Link &  &  & 10527 &  \\ \cline{3-3} \cline{6-6}
 &  & WithingsCam &  &  & 14088 &  \\ \hline
\begin{tabular}[c]{@{}c@{}}Set III\\ (Others)\end{tabular} & \begin{tabular}[c]{@{}l@{}}Own \\ Laboratory\end{tabular} & Others & 22.57 & 3,11,94,288 & 90669 & 80849 \\ \hline
\end{tabular}
\end{table*}

For the experiments in this paper, we have considered two large sets of data collected from two different sources. 
One set of data is collected from our own laboratory that includes six commercially available IoT cameras, and the other set of data is open-sourced data by UNSW \cite{Arunan2018IEEE}. A summary of the data sets is given in Table \ref{tab:camera-data-set}.

In the former case, the Internet connectivity of the IoT cameras is controlled by the Wi-Fi credential of a customized Wi-Fi router, created using a Raspberry Pi Model B; hence forth simply called as router. 
To create ground truth, network traffic is collected when only a specific IoT camera is active. 
\texttt{Ethernet} interface of the router is used to sniffs network traffic, using \texttt{tshark}. 
In this case, two broad datasets are collected, Set I and Set III.

In Set I (Named as BITSPHC), six commercially off-the-self IoT cameras (Ezviz, D3D, V380 Spy Bulb, Netatmo, Canary, and Alarm Spy Clock) are used to collect a sum total of about 14.96GB of pcap files. 
The cameras are accessed remotely by their companion mobile applications installed on a Samsung Tab and the Tab is connected to a different network for Internet connectivity. 
Accessing an IoT camera involves turning on all of its features, such as tilt, pan, 360-degree movement, and audio on/off mode, while also gathering traffic. 
The traffic from all these six IoT cameras is assigned a broad label as \texttt{IoTCam}, and the traffic from individual IoT camera is labeled as \texttt{Ezviz}, \texttt{D3D}, \texttt{Spy Bulb}, \texttt{Netatmo}, \texttt{Canary} and \texttt{Spy Clock} to indicate the specific IoT camera. 

The open-sourced data from UNSW dataset \cite{Arunan2018IEEE} is in Set II (called as UNSW), and this set consists of network traffic, i.e., pcap files, from five different IoT cameras. 
The traffic of the data set is filtered based on the IP addresses and description provided in the \enquote{read me} file of the dataset. To distinguish the cameras, the traffic of individual IoT cameras is labeled as \texttt{Dropcam}, \texttt{Netatmo}, \texttt{SamsungSmartCam}, \texttt{TPlink} and \texttt{Withings}. 
This dataset turns out to have 1.49GB, and it includes five IoT cameras, where only one camera (i.e., Netatmo) is in common with Set I. This dataset can allow us to extend the applicability of our proposed system into a diverse of IoT cameras.

In Set III (called as Others), we consider the network traffic of other networked applications, including but not limited to online audio/video conferencing (e.g., Meet, Skype, Zoom, etc.), video sharing (e.g., Amazon Prime, YouTube, etc.), and other applications (e.g. reading text, newspapers, etc.) that do not produce any audio/video applications. Overall, this dataset, having a size of about 22GB, represents a dataset that does not belong to any IoT camera.
This dataset forms the basis of one class classification, and Set I and Set II are considered as representative of zero-day IoT cameras.

The table \ref{tab:hyper_oc} outlines the hyperparameter settings for the one-class classifiers. 
Each model applies a different strategy for detecting the anomalies. 
OCSVM uses a tight boundary around the benign data points in a transformed feature space, typically using an Radial Basis Function (RBF) kernel, i.e., $K(x,y) = \nu * exp(-\gamma * ||x-y||^2)$, where $\nu$ and $\gamma$ are hyperparameters. 
With a very low $\nu$ \textit{nu} value (0.001), the model assumes almost all training data are normal, enforcing a strict enclosure. 
A high $\gamma$ \textit{gamma} value (0.999) makes the boundary highly sensitive to subtle variations in data, enabling fine-grained anomaly detection. 
This configuration is well-suited for detecting the anomalies for the consideration of zero-day IoT cameras.
In SGDOCSVM, with \textit{$\nu$ = 0.03}, it allows up to 3\% anomalies, offering more flexibility than OCSVM. 
A smaller \textit{learning rate $\eta$0 = 0.0001} ensures stable and gradual convergence. 
This setting provides a balance between anomaly sensitivity and training efficiency. IF model randomly partitions the feature space, it isolates data points, with anomalies requiring fewer splits due to their deviation from normal patterns. 
We use \textit{contamination level} as 0.1, where the model is tuned to be highly sensitive, prioritizing the detection of subtle anomalies at the cost of potential false positives. 
Its efficiency and ability to handle high-dimensional data can support real-time anomaly detection in resource-constrained IoT environments.
We used DeepSVDD, to learn a compact representation of normal IoT camera behavior by mapping input data into a latent feature space, where a minimal-volume hypersphere is optimized to enclose the majority of benign instances. 
The network architecture consists of two fully connected hidden layers, each with \textit{512 neurons}, having \textit{ReLU} activation function, to capture non-linear patterns. 
The output layer reduces the representation to a \textit{latent} dimension of 8, enabling tight modeling of normal traffic. 
This setup can particularly be suited for zero-day anomaly detection, as the novel or abnormal behaviors — unseen during training —are projected far from the center of the hypersphere. 

\begin{table}[h]
\centering
\caption{Hyperparameters Used in the one class classifications}
\label{tab:hyper_oc}
\begin{tabular}{ccc}
\hline
Model Name & Parameter Name & Parameter Values \\ \hline \hline
\multirow{2}{*}{OCSVM} & nu & 0.001 \\ \cline{2-3} 
 & gamma & 0.999 \\ \hline
\multirow{2}{*}{SGDOCSVM} & nu & 0.03 \\ \cline{2-3} 
 & eta0 & 0.0001 \\ \hline
IF & contamination & 0.1 \\ \hline
\multirow{7}{*}{DeepSVDD} & Input Dimension & 10 \\ \cline{2-3} 
 & Hidden Dimension & 512 \\ \cline{2-3} 
 & Latent Dimension & 8 \\ \cline{2-3} 
 & Learning Rate & 0.0001 \\ \cline{2-3} 
 & Epochs & 150 \\ \cline{2-3} 
 & Optimiser & Adam \\ \cline{2-3} 
 & Threshold & 95 \\ \cline{2-3} \hline
\end{tabular}
\end{table}

\begin{figure*}[h!]
    \begin{subfigure}{.33\linewidth}
    \centering
    \includegraphics[width=\linewidth,height=4.5cm]{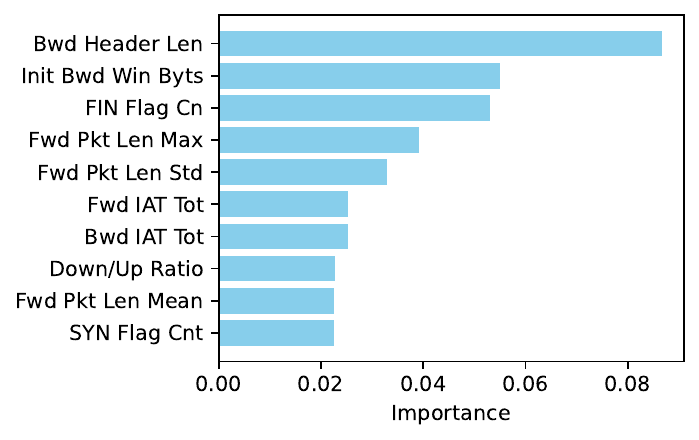}  
    \caption{Feature Importance}
\end{subfigure}
\begin{subfigure}{.33\linewidth}
    \centering
    \includegraphics[width=\linewidth,height=4.5cm]{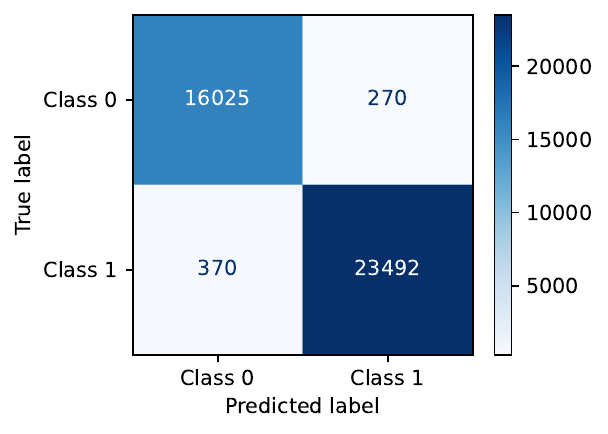}  
    \caption{Confusion Matrix} 
\end{subfigure}
\begin{subfigure}{.33\linewidth}
    \centering
    \includegraphics[width=\linewidth,height=4.5cm]{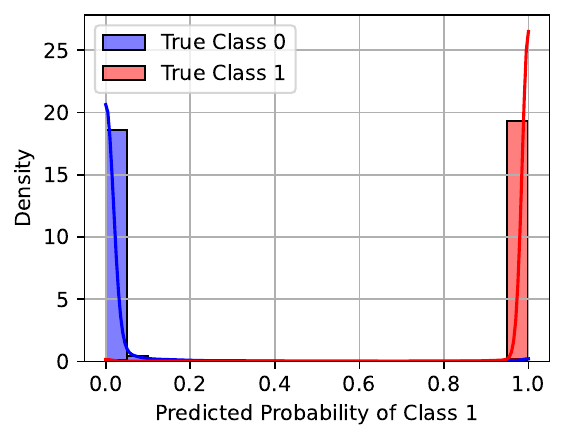}  
    \caption{Prediction Probabilities}
\end{subfigure}
\caption{Performance of ExtraTree model in feature selection having two classes as \textit{Others (Class 0)} and \textit{IoTCam (Class 1)}.}
\label{fig:featSelection_ET}
\end{figure*}

\section{Detecting Zero-Day IoT (Spy) Cameras}
\label{sec:detect-zero-day}
We perform an extended set of experiments to detect zero-day IoT cameras. 
In zCamInspector, zCamDetector is basically a one-class classifier that detects any zero-day IoT camera. 
A Zero-day IoT camera is an IoT camera whose network traffic has not been used while training the zCamDetector. 
zCamDetector can uses any of the four models, namely \texttt{OCSVM}, \texttt{SGDOCSVM}, \texttt{IF} and \texttt{DeepSVDD}.
In general, whenever any of the these models detects an outlier, it is considered as a zero-day IoT camera in this paper.

SGDOCSVM is a variant over OCSVM \cite{yang2021ocsvm}, that can achieves a better precision in outlier detection due an optimization applied on the objective function and iteratively updating the model parameters using subsets of the data \cite{SGDOCSVM2024Sciencedirect}. 
SGDOCSVM significantly reduces the computation time without sacrificing accuracy much when streaming data is used for classification \cite{liu2020IoTJ}. 
Both OCSVM and SGDOCSVM achieve better model fitting and reduce prediction times of outlier \cite{SGDOCSVM2024Sciencedirect}.
IF detects anomalies by randomly selecting features and splitting their values in fewer steps. 
It is a scalable one-class classifier and ideal for high-dimensional data with limited labeled samples \cite{Isolation2008IEEE, tan2022Elsevier, verma2023zeroIEEE}.
DeepSVDD is an advanced anomaly detection technique that leverages deep learning to model the boundary of normal data in a high-dimensional space. 
By combining the power of Support Vector Data Description (SVDD) and neural networks, it learns a decision boundary that efficiently separates normal data from anomalies, even in complex and high-dimensional data such as images, text, or tweets. 
DeepSVDD outperforms traditional anomaly detection methods, particularly when the data is not linearly separable, \cite{ruff2018dsvdd}, \cite{sohn2021deepOCclass} 
Thus, we consider all of these four models in our study for the detection of zero-day IoT cameras.

We consider a total of 11 IoT cameras, combinely present in Set I and Set II, for this set of experiments. 
Before applying any of the models, we investigate the features by performing clustering using the Principal Component Analysis (PCA) and Gaussian Mixture Model (GMM) to understand the potential of the features in detecting zero-day IoT cameras. 
PCA is a powerful mechanism to reduce the feature dimension and uncover any anomaly in the dataset by monitoring the reconstruction errors.  
On the other hand, GMM is a probabilistic model that assumes data is generated from a mixture of several Gaussian distributions, each is a component in GMM with its own mean $\mu_i$ and variance $\sigma_i$. 
It is expected that each component represents a cluster in the given data.  
The model assigns probabilities to data points based on their likelihood of belongingness to each of the clusters. 
Thus, GMM can be used to detect anomalies by thresholding the probabilities for each sample to belong to any of the clusters. 

\begin{figure*}[hbtp]
    \begin{subfigure}{.33\linewidth}
    \centering
    \includegraphics[width=\linewidth]{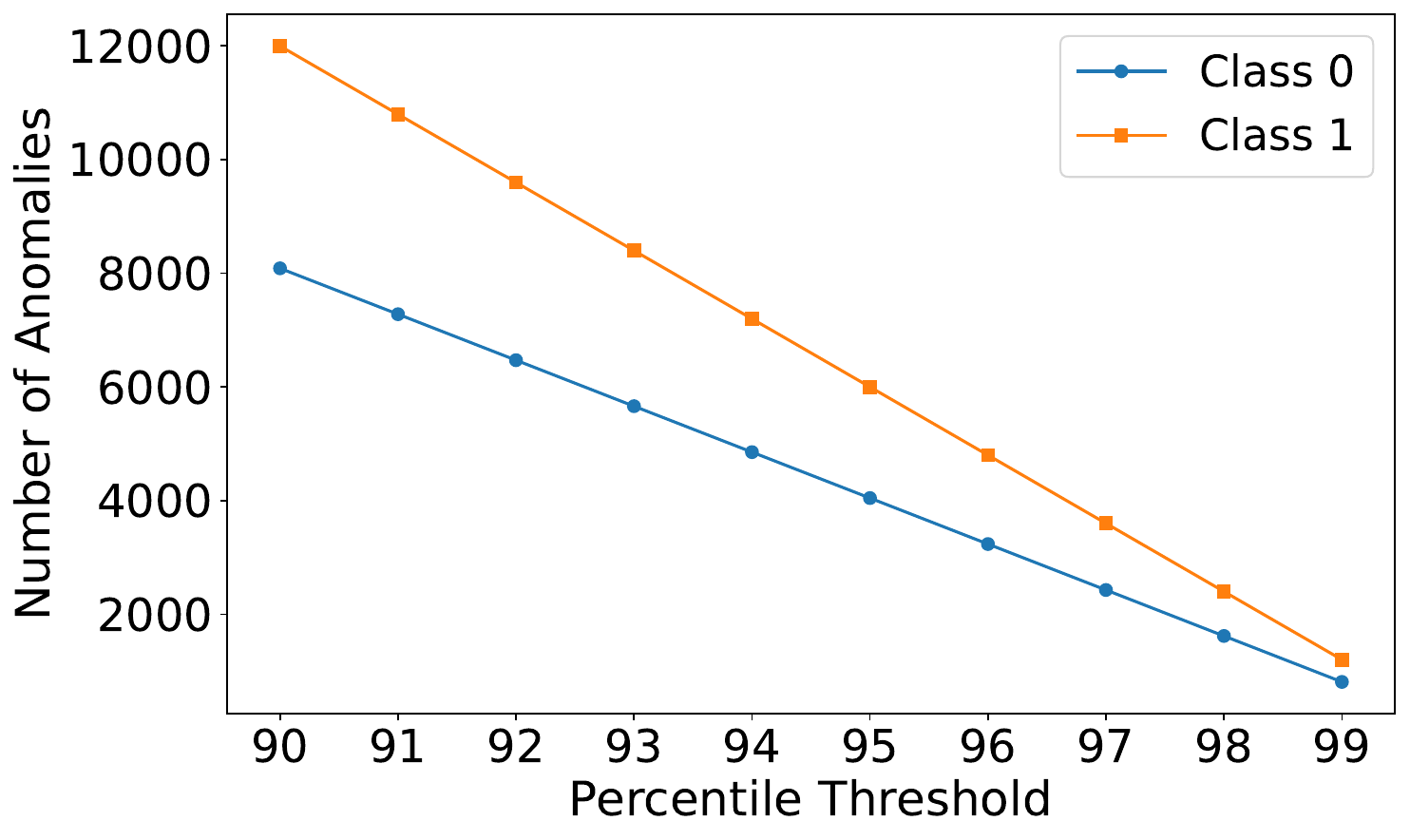}  
    \caption{Number of anomalies per threshold in reconstruction error}
\end{subfigure}
\begin{subfigure}{.33\linewidth}
    \centering
    \includegraphics[width=\linewidth]{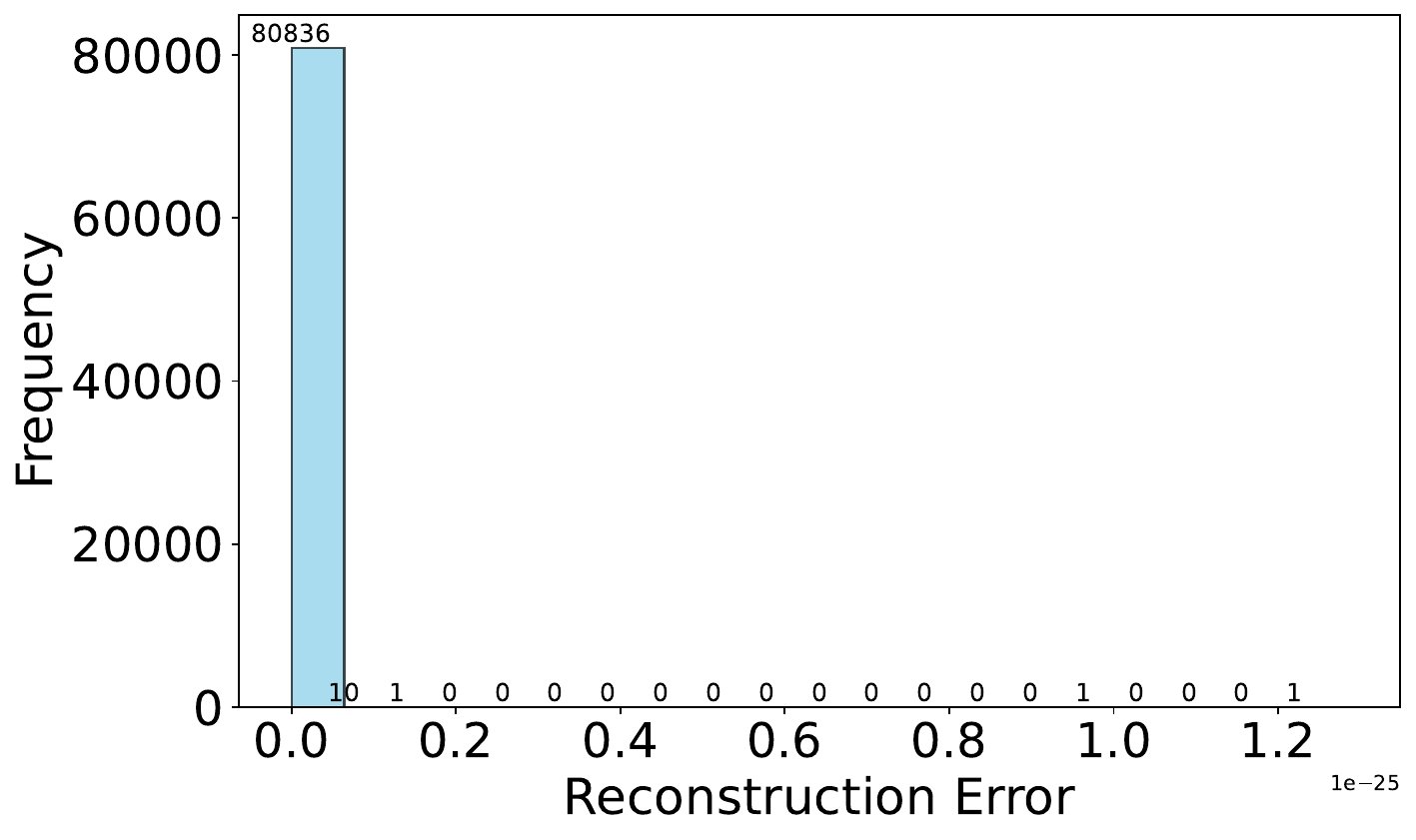}  
    \caption{Distribution of reconstruction error in Class 0} 
\end{subfigure}
\begin{subfigure}{.33\linewidth}
    \centering
    \includegraphics[width=\linewidth]{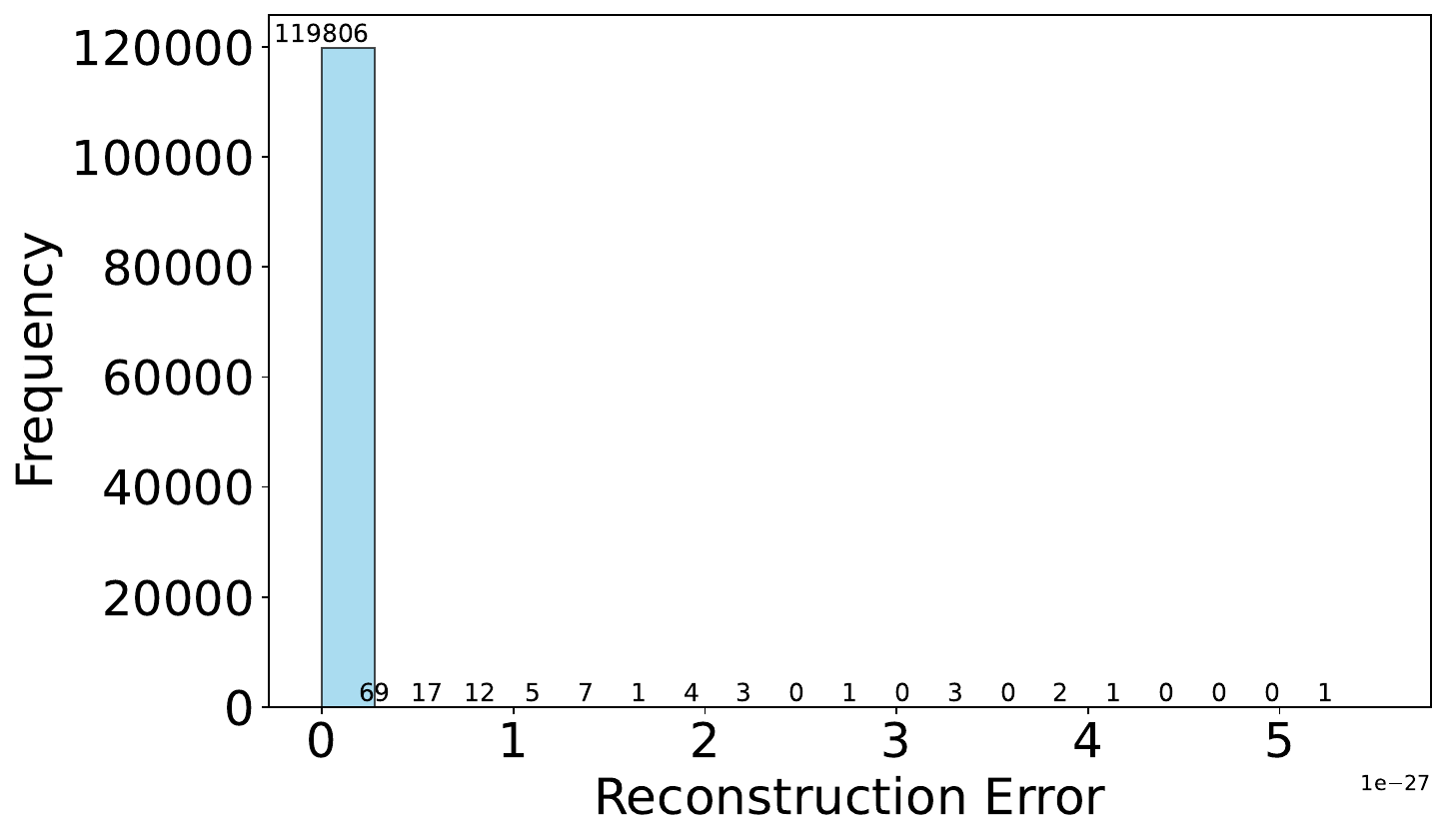}  
    \caption{Distribution of Reconstruction Error in Class 1}
    \end{subfigure}
\caption{Anomaly detection using PCA in Others (Class 0) and IoTCam (Class 1).}
\label{fig:pca}
\end{figure*}

\subsection{Results of Feature Selection using ExtraTree Model}

We use only the top 10 features based on the ExtraTree classifier for this classification. 
ExtraTree model is basically a supervised classifier that has shown a strong ability for feature selection in a variety of applications, particularly due to the built-in feature importance calculation, e.g., using gini index or entropy, and the avoidance of feature scaling.   
In this paper, we consider two classes,  "Others", i.e., Set III, and  "IoTCam", i.e., the combined set of Set I and Set II, for selecting the features using this classifier.  

Figure \ref{fig:featSelection_ET}(a) shows the top 10 most important features along with their feature important score. 
Figure \ref{fig:featSelection_ET}(b) shows the confusion matrix of the two-class classification, and it turns out that the misclassification rate is as low as 0.0025\%. 
Figure \ref{fig:featSelection_ET}(c) shows the density of the prediction probabilities for the two classes. 
The probabilities closer to zero and one indicate the class of \textit{Others} and \textit{IoTCam} respectively; and, the decision boundary can be clearly visible. 

Once the top 10 features are obtained, we go on to detect outliers by applying conventional clustering via PCA and GMM. 
Let us denote by $F_k$ the set of flows that are known to the classifier and $F_z$ the set of flows that are not known to the classifier. 
For example, if the flows in Set III are known, then  $F_k = "Others"$, and if none of the flows of 11 IoT cameras are known, then  $F_z = Set I \cup Set II$. 
In other words, $F_z$ denotes the traffic of zero-day IoT cameras.
Thus, assuming that a network administrator has not seen any traffic from any of the IoT Cameras, A model in zCamDetector is trained only with Set III. 
The administrator is interested in knowing whether an IoT camera connects to the network under surveillance.

\subsection{PCA to Detect Outliers In $F_k$ and $F_z$}
After scaling using \textit{standard scaler}, we first fit both $F_k = Set III$ and $F_z = Set I \cup Set II$ and then reverse transform to reconstruct the original data. Then, by applying a threshold of 95 percentile on the reconstruction error, we consider the samples having errors beyond this threshold as outliers in both $F_k$ and $F_z$. 
Figure \ref{fig:pca}(a) shows the ground truth samples from both $F_k$ and $F_z$, and Figure \ref{fig:pca}(b) shows the PCA induced outliers in each of the datasets separately. 
It turns out that 5.81\% and 4.07\% of samples in $F_k$ and $F_z$ are outliers. 
Figure \ref{fig:pca}(c) shows the reconstruction errors when 95 percentile is used as the threshold. 
No surprise that this threshold can be tuned to control the number of outliers.

\subsection{GMM PCA to Detect Outliers In $F_k$ and $F_z$} 
\begin{figure}[h]
\centering
\begin{subfigure}{0.45\textwidth}
    \centering
    \includegraphics[width=\linewidth]{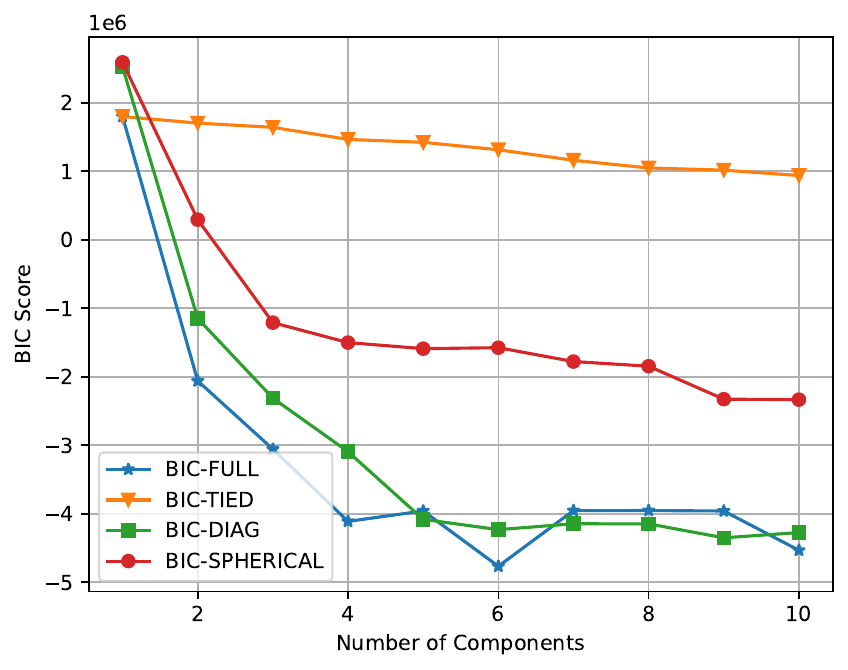}  
    \caption{Best Component using BIC on Others Dataset}
\end{subfigure}
\begin{subfigure}{0.45\textwidth}
    \centering
    \includegraphics[width=\linewidth]{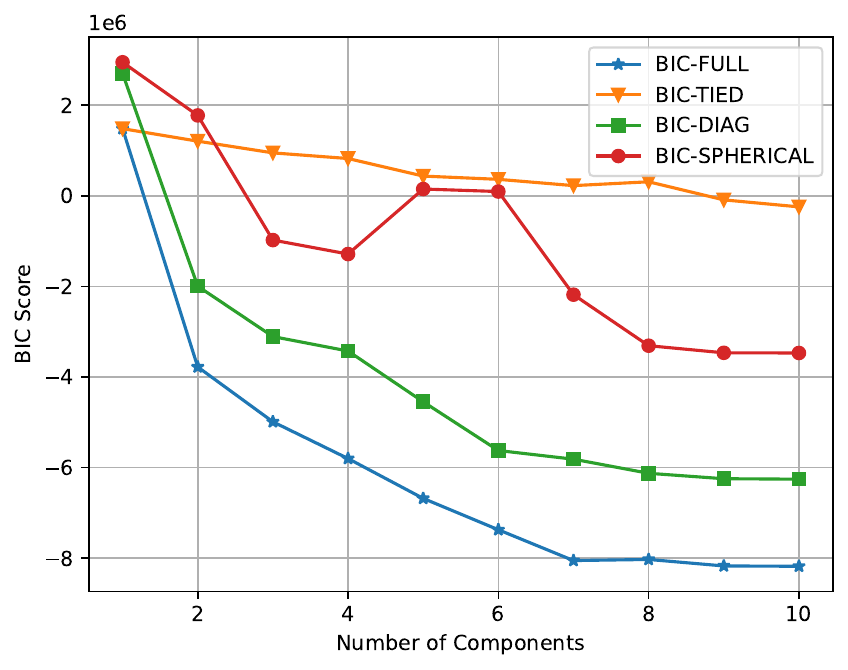}  
    \caption{Best Component using BIC on IoTCam Dataset}
\end{subfigure}
\caption{Best Component using BIC and GMM}
\label{fig:BIC_gmm}
\end{figure}

\begin{figure*}[h]
\centering
    \begin{subfigure}{.32\textwidth}
    \centering
    \includegraphics[width=\linewidth]{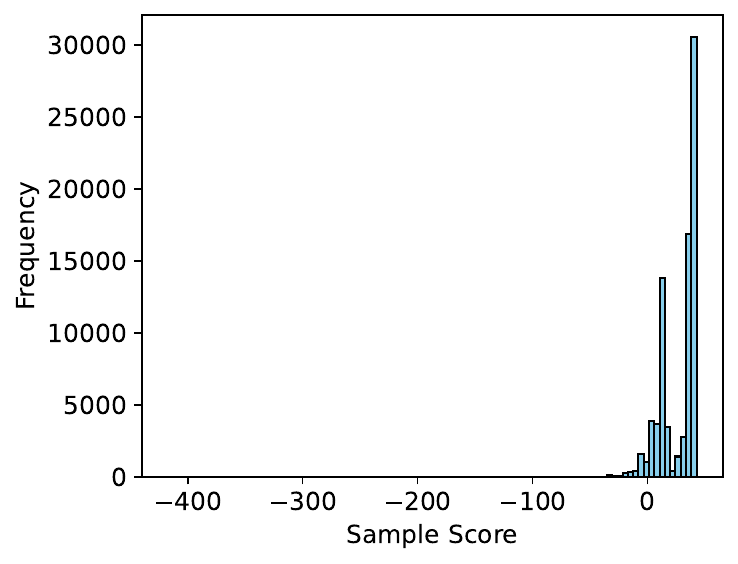}  
    \caption{Sample Score using GMM for Others (Class 0) }
\end{subfigure}
\begin{subfigure}{.32\textwidth}
    \centering
    \includegraphics[width=\linewidth]{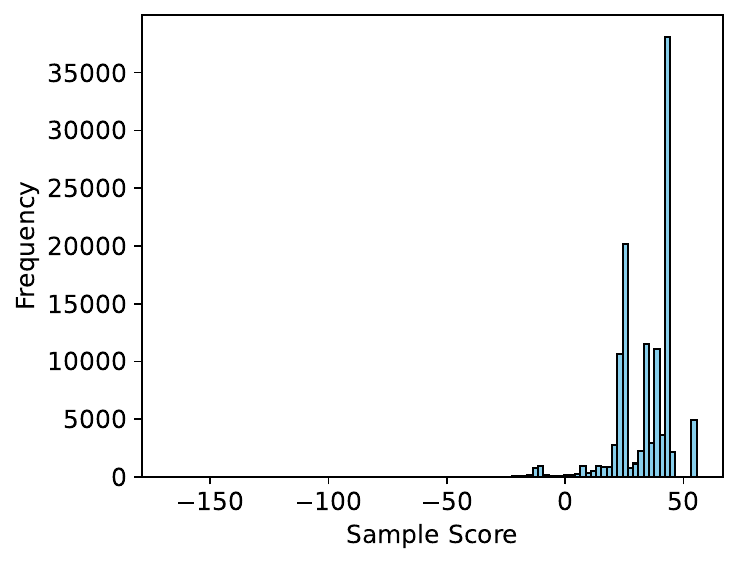}  
    \caption{Sample Score using GMM for IoTCam (Class 1) } 
\end{subfigure}
\begin{subfigure}{.32\textwidth}
    \centering
    \includegraphics[width=\linewidth]{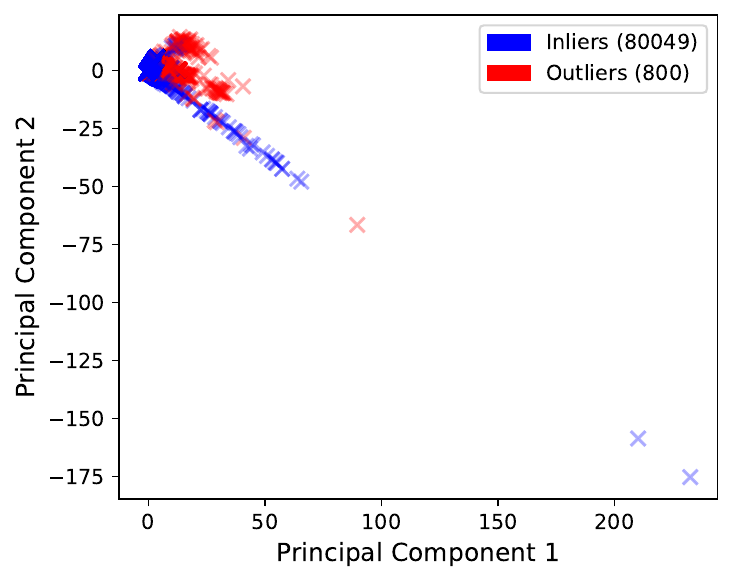}  
    \caption{Outliers in Others }
    \end{subfigure}
\begin{subfigure}{.32\textwidth}
    \centering
    \includegraphics[width=\linewidth]{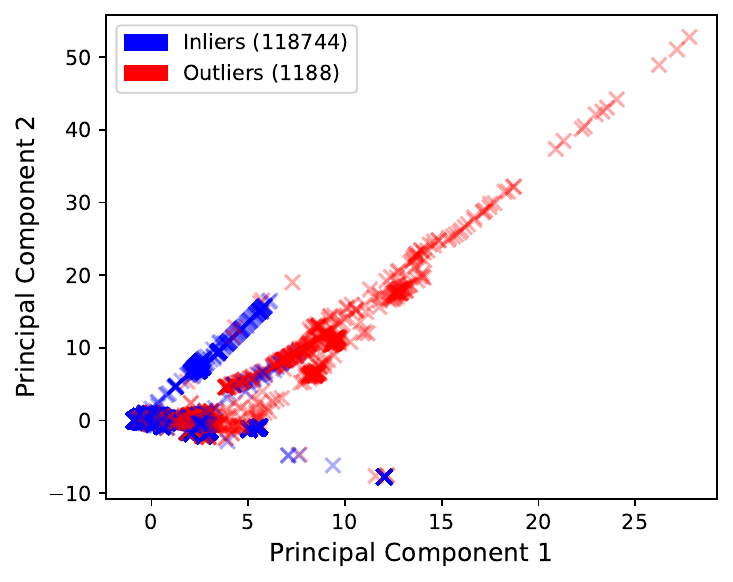}  
    \caption{Outliers in IoTCam }
    \end{subfigure}
\begin{subfigure}{.32\textwidth}
    \centering
    \includegraphics[width=\columnwidth]{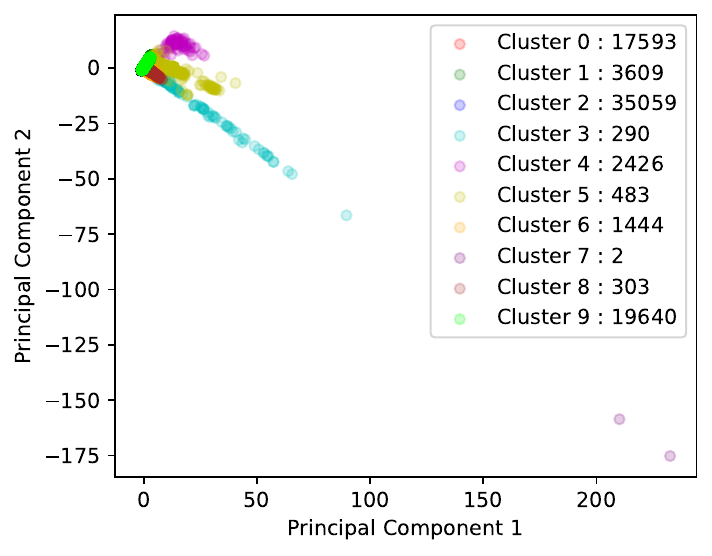}  
    \caption{Others Dataset}
\end{subfigure}
\begin{subfigure}{.32\textwidth}
    \centering
    \includegraphics[width=\columnwidth]{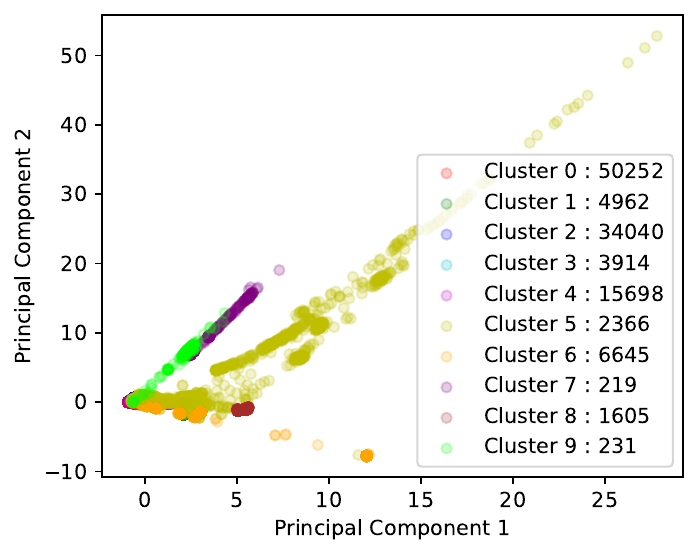}  
    \caption{IoTCam Dataset}
\end{subfigure}
\caption{Anomaly detection using GMM in Others (Class 0) and IoTCam (Class 1).}
\label{fig:gmm}
\end{figure*}

After scaling using \textit{standard scaler}, we fit GMM by varying the number of components from 1 to 20 on $F_k = Set III$ and the \texttt{variance} as any of \texttt{spherical, tied, diag and full}. 
We use the Bayesian Information Criterion (BIC) to find the minimum number of components in GMM that fits the data well; a lower BIC value indicates a better fit of GMM with the data.

Figure \ref{fig:gmm}(a) and (b) shows both the data sets, i.e., $F_k$ and $F_z$, with sample score using GMM. Figure \ref{fig:gmm}(c) and (d) shows PCA-reduced two features with the inliers and outliers when 95 percentile is considered as threshold, resulting roughly 5\% of the samples in Set III as outliers. Applying on $F_z$, the same GMM finds 5\% (1188 out of 119932) of outliers when all 11 IoT cameras are zero-day. The datasets have significant overlap, indicating that detection of zero-day IoT camera in $F_z$ may not be trivial. 
Figure \ref{fig:BIC_gmm}(a) and (b) shows the BIC values for the GMMs with different number of components. It turns out that GMM with 9 components and \emph{diag} variance best fits $F_k$ and $F_z$, i.e., GMM requires  a minimum of 9 clusters with \emph{diag} variance to fit $F_k$ and $F_z$ dataset(shown in Figure \ref{fig:gmm}(e) and (f)).

\subsection{Comparison of the Outliers by PCA and GMM}

\begin{table}[hbtp]
\centering
\caption{Comparison of the Outliers by PCA and GMM at 99\% percentile. \#comps and Cov. type indicate the number of components and the covariance type respectively, and seed is the seed for random sampling.}
\label{tab:outlier_PCA_GMM}
\begin{tabular}{|c|cc|ccc|}
\hline
\multirow{2}{*}{Method} & \multicolumn{2}{c|}{Datasets} & \multicolumn{3}{c|}{Parameters} \\ \cline{2-6} 
 & \multicolumn{1}{c|}{Others} & IoTCam & \multicolumn{1}{c|}{\#comps} & \multicolumn{1}{c|}{cov. type} & seed \\ \hline
PCA & \multicolumn{1}{c|}{809} & 1200 & \multicolumn{1}{c|}{10} & \multicolumn{1}{c|}{---} & 42 \\ \hline
GMM & \multicolumn{1}{c|}{800} & 1188 & \multicolumn{1}{c|}{10} & \multicolumn{1}{c|}{Full} & 42 \\ \hline
\end{tabular}
\end{table}

Table \ref{tab:outlier_PCA_GMM} provides a comparative analysis of outlier detection using PCA and GMM at $99^{th}$ percentile as threshold. The evaluation is conducted on two datasets, Others and IoTCam, using a consistent configuration of 10 components and a fixed random state of 42. 
PCA has identified 809 and 1200 outliers in the Others and IoTCam datasets respectively, whereas GMM has detected slightly fewer outliers—800 and 1188 respectively. 
GMM is configured with a full covariance type, whereas PCA did not require this specification.
The results indicate that GMM exhibits a more conservative and probabilistically grounded approach to outlier detection, potentially minimizing false positives. Consequently, GMM is more suitable than PCA for accurate anomaly detection in high-dimensional IoT camera datasets.


\subsection{Zero-Day IoT Camera using One-Class Models}

\begin{figure*}[hbtp]
\centering  
\includegraphics[width=\linewidth]{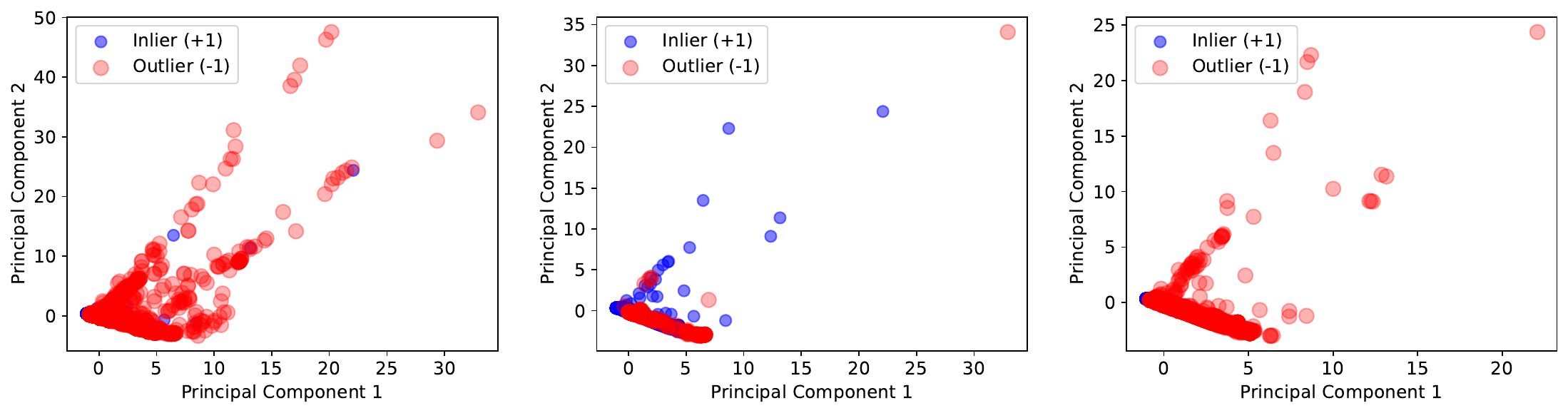}  
    \caption{Plot of the training samples from "Others" class, where each of the three models, i.e., ocsvm, sgdocsvm and iso-forest, is trained using training samples of Others.}
    \label{fig:trainOthers_PredictTrainOthers}
\end{figure*}     

\begin{figure*}[hbtp]
\centering  
\includegraphics[width=\linewidth]{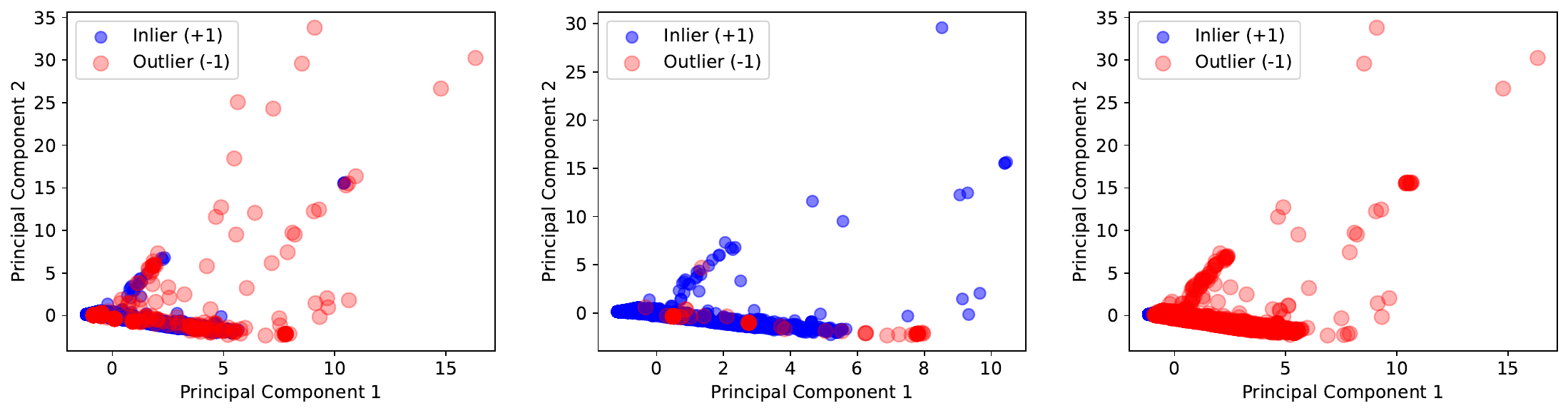}  
    \caption{Plot of test samples of "Others" class, where each of the three models, i.e., ocsvm, sgdocsvm and iso-forest, is trained on the training samples of "Others".}
    \label{fig:trainOthers_PredictontestOthers}
\end{figure*}

\begin{figure*}[hbtp]
\centering  
\includegraphics[width=\linewidth]{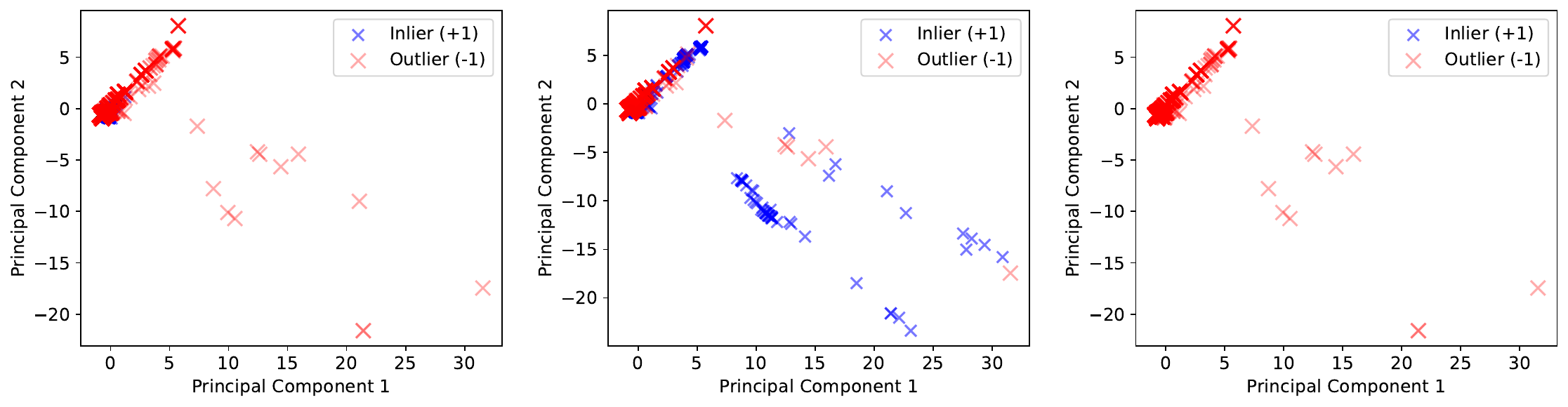}  
    \caption{Plot of the test samples from "IoTCam" dataset, where each of the three models, i.e., ocsvm, sgdocsvm and iso-forest, is trained using training samples of Others.}
    \label{fig:trainOthers_predictontestIoTCam}
\end{figure*} 

 

\begin{figure*}[hbtp]
\begin{subfigure}{.32\textwidth}
    \centering
    \includegraphics[width=\columnwidth]{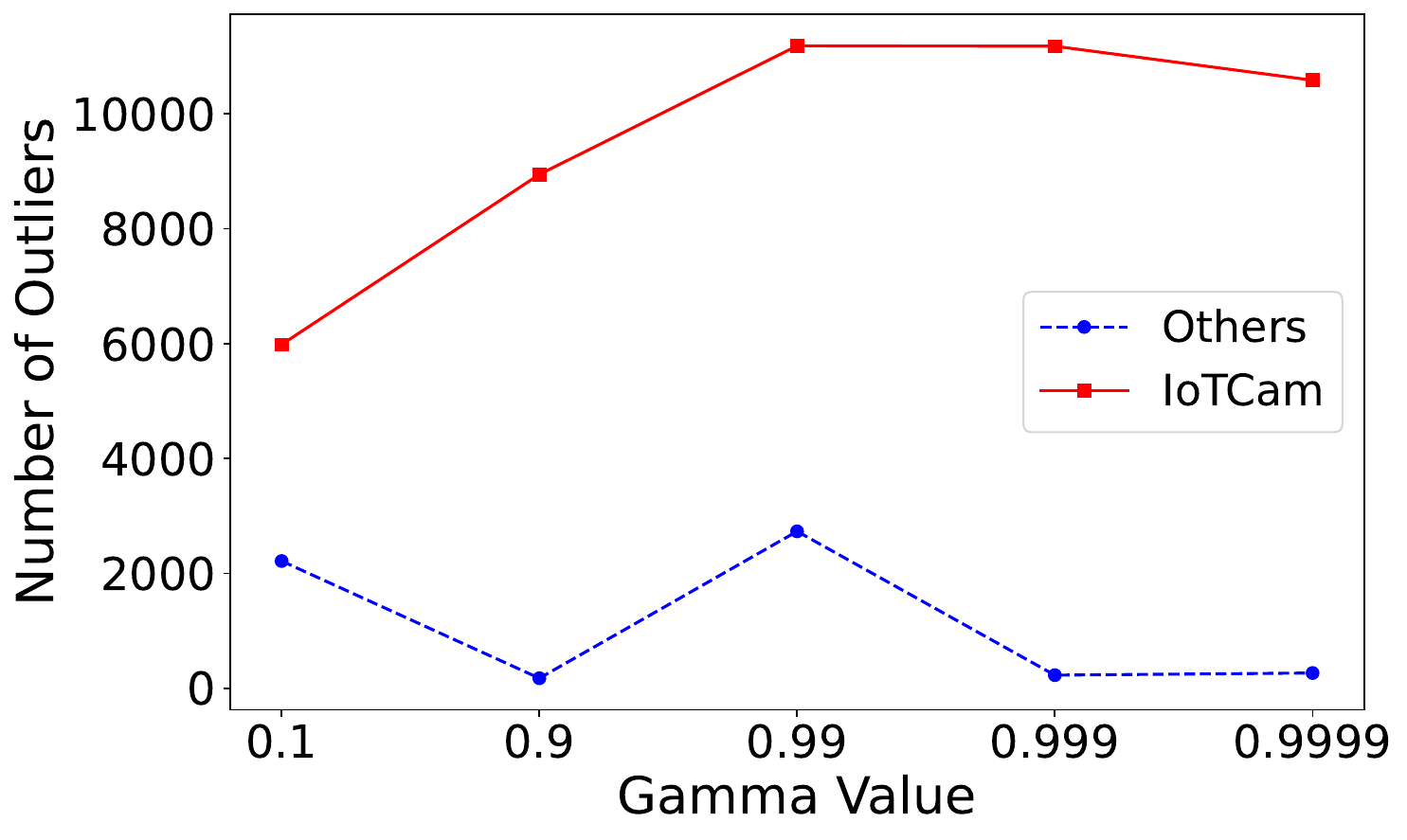}  
    \caption{Outliers in OCSVM}
\end{subfigure}
\begin{subfigure}{.32\textwidth}
    \centering
    \includegraphics[width=\columnwidth]{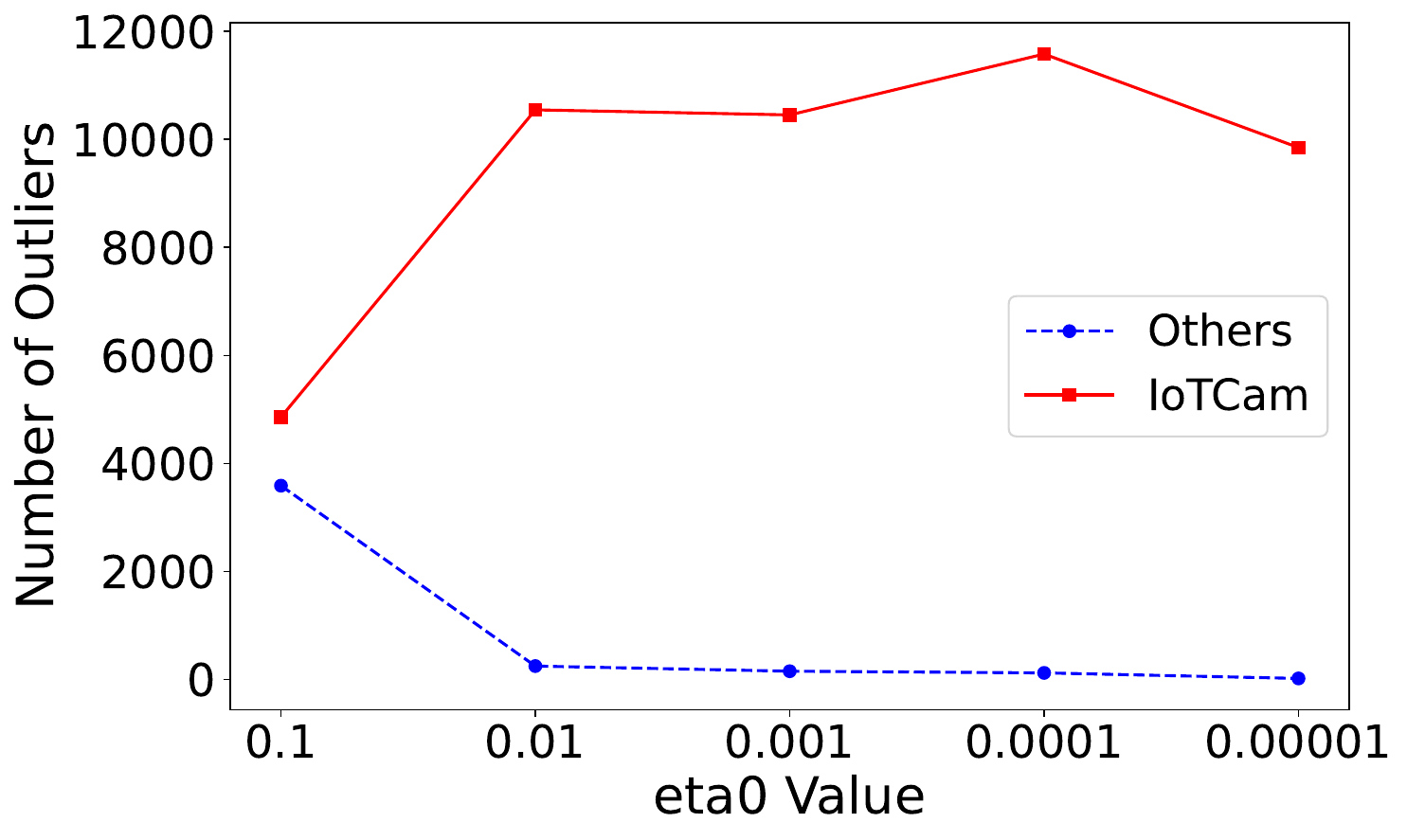}  
    \caption{Outliers in SGDOCSVM}
\end{subfigure}
\begin{subfigure}{.32\textwidth}
    \centering
    \includegraphics[width=\columnwidth]{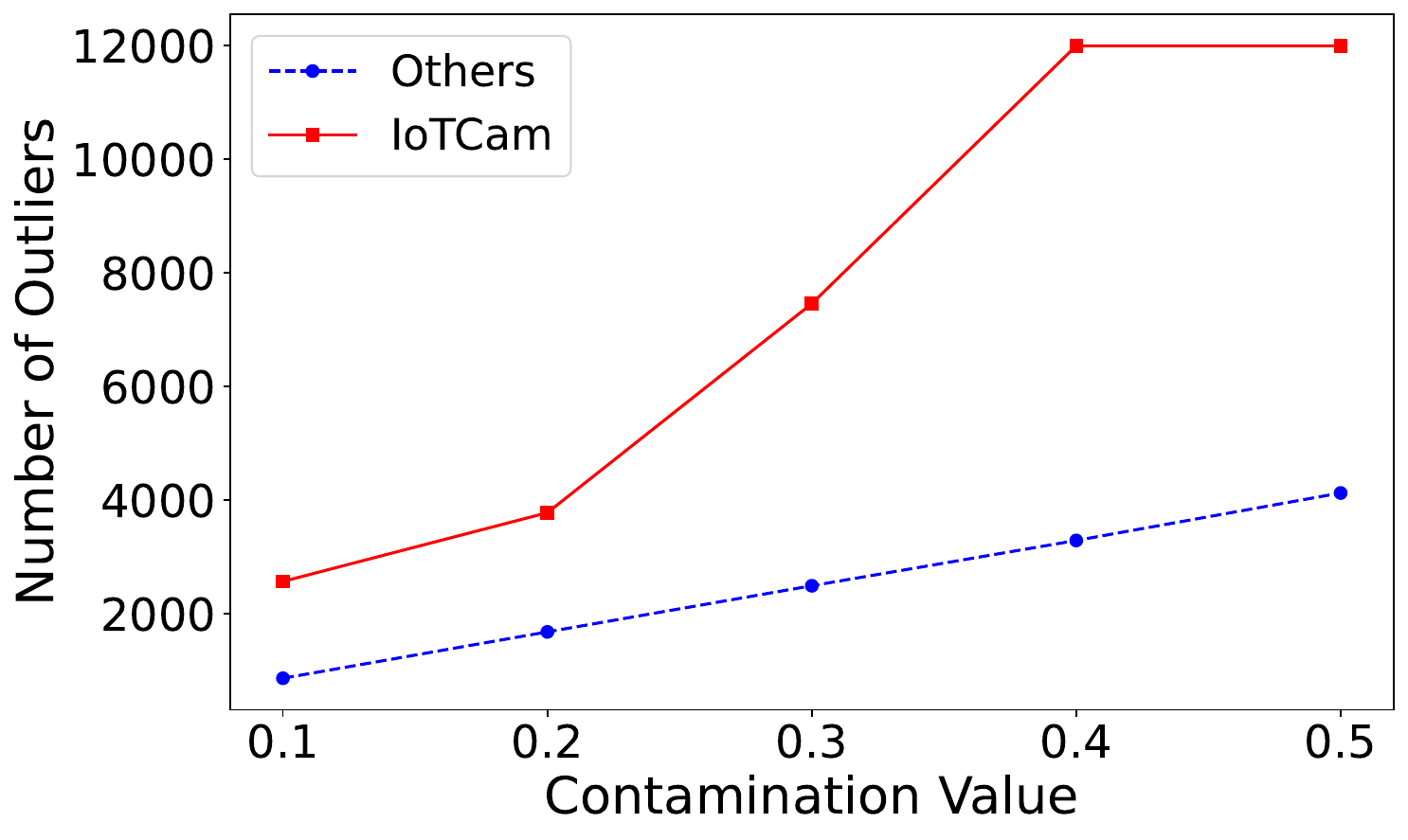}  
    \caption{Outliers in Isolation forest}
\end{subfigure}
\caption{Outlier detection by tuning different parameters using one class classification models.}
\label{fig:outlierdetection_parametertuning}
\end{figure*}

We initiated our investigation by training 90\% data in Set III, i.e., Others dataset, using the top 10 features, and kept the remaining 10\% for testing. Fig. \ref{fig:trainOthers_PredictTrainOthers} shows inliers and outliers plot of the training samples of Others class, where each of the three models i.e. OCSVM, SGDOCSVM, IF is trained using training samples of the same dataset i.e. Others. We use the same trained model and shows the inliers and outliers detected on test samples of the Others class shown in Fig. \ref{fig:trainOthers_PredictontestOthers}. 
Fig. \ref{fig:trainOthers_predictontestIoTCam} shows both the outliers and inliers in the test set of IoTCam when trained using trained samples of the Others. 
We shows the number of outliers in OCSVM, SGDOCSVM and IF by tuning different hyperparameter in Fig. \ref{fig:outlierdetection_parametertuning} indicating maximum outliers in IoTCam class. 
We show mean and standard deviation in Table \ref{tab:mean_std_diffclassifiers} where our analysis shows DeepSVDD having minimum standard deviation in training (Set III) and while testing in Set I, Set II and its combination (Set I and Set II).

\begin{table*}[hbtp]
\centering
\caption{Mean and Standard Deviation (Std. Dev.) during testing on Set I, Set II, and Combine (Set I and Set II) using OneClassSVM, SGDOneClassSVM, Isolation Forest and DeepSVDD, where the models has been trained only using \enquote{Others} dataset.}
\label{tab:mean_std_diffclassifiers}
\begin{tabular}{ccccccccccc}
\hline
\multirow{3}{*}{\begin{tabular}[c]{@{}c@{}}{\diagbox[width=10em]{Classifiers}{Datasets}}\end{tabular}} 
\multirow{3}{*}{} & \multicolumn{4}{c}{Set III} & \multicolumn{2}{c}{Set I} & \multicolumn{2}{c}{Set II} & \multicolumn{2}{c}{Set I \& Set II} \\ \cline{2-11} 
 & \multicolumn{2}{c}{Training} & \multicolumn{2}{c}{Testing} & \multicolumn{2}{c}{Testing} & \multicolumn{2}{c}{Testing} & \multicolumn{2}{c}{Testing} \\ \cline{2-11} 
 & Mean & Std. Dev. & Mean & Std. Dev. & Mean & Std. Dev. & Mean & Std.Dev. & Mean & Std. Dev \\ \hline \hline
OneClassSVM & 90.44 & 0.49 & 90.6 & 3.29 & 29.03 & 1.04 & 20.5 & 0.74 & 27.2 & 0.72 \\ \hline
SGDOneClassSVM & 90.58 & 7.81 & 83.75 & 11.08 & 18.8 & 12.78 & 17.13 & 16.12 & 19.7 & 13.97 \\ \hline
Isolation Forest & 90.08 & 0.13 & 91 & 3.76 & 20.55 & 0.19 & 10.15 & 0.39 & 14.35 & 0.31 \\ \hline
DeepSVDD & 99.45 & 0.002 & 99.35 & 0.002 & 99.54 & 0.19 &  99.75 & 0.004 & 98.17 & 0.05 \\ \hline
\end{tabular}
\end{table*}

\subsection{All But one Zero Day}
Now we apply the classifiers as follows. 
At first, a model is trained with $F_n$, where $F_n$ contains the flows that belong to only one IoT camera. And the flows of the remaining 10 IoT cameras belong to $F_i$. Thus making a testing set having a much higher amount of data points compared to training set.
This setting represents a scenario where the administrator has knowledge about only one camera and not about the rest. 
Thus, we have performed a set of 11 experiments, i.e., 11 trained models, with each having flows of one out of 11 cameras. 
For example, in the first experiment, we have considered the flows in D3D camera belong to $F_n$ to train one OneClassSVM model, and the remaining 10 cameras belong to $F_i$, i.e., $F_i$ contains flows of Ezviz, Netatmo, SpyClock, Canary, SpyBulb, Dropcam, Netatmo, Samsung SmartCam, TP-Link, WithingsCam. 
Table \ref{tab:summary_OneClassSVMipcam} compares One-Class SVM, SGDOneClassSVM, Isolation Forest, and DeepSVDD based on accuracy and testing performance (inliers vs. outliers). 

In general, while the accuracy in training turns out to be more than 80\%, the accuracy in testing is less than 30\%. In other words, a vast majority of the unseen camera flows were detected as outliers in DeepSVDD.
The performance evaluation of four anomaly detection models—DeepSVDD, Isolation Forest, SGDOneClassSVM, and One-Class SVM—for detecting zero-day IoT (Spy) cameras reveals clear distinctions in their strengths and limitations.
We observes \textbf{DeepSVDD} emerges as the most robust and consistent model, achieving the highest \textbf{testing accuracy in 7 out of the 11 test scenarios}. It demonstrates superior capability in separating outliers, especially in complex environments, as evidenced by its outstanding performance on \textit{Spy Bulb} (99.06\% testing accuracy). This consistency makes it well-suited for generalizing across diverse IoT camera types. However, its performance slightly dips when tested with \textit{Netatmo} (52.16\%) and \textit{D3D} (52.78\%), indicating potential sensitivity to certain device-specific characteristics or limited feature generalization in these cases.
\textbf{Isolation Forest} ranks as the second-best performer overall. It exhibits high testing accuracy in several instances, including excellent results for \textit{TP-Link} and \textit{DropCam}. Nonetheless, it suffers from notable degradation in performance for specific devices like \textit{Netatmo} (28.53\%) and \textit{D3D} (53.73\%), suggesting that while it performs well under many conditions, it lacks robustness in the presence of certain data distributions or device types. \textbf{SGDOneClassSVM} shows moderate effectiveness, delivering reasonable accuracy in a number of cases and outperforming traditional One-Class SVM in some scenarios with limited data. Despite these improvements, the model remains inconsistent and unstable, with poor accuracy results on devices such as \textit{SpyClock}, \textit{Canary}, and \textit{Netatmo}, which limits its reliability in critical zero-day attack detection settings. Lastly, \textbf{One-ClassSVM} consistently demonstrates the weakest performance among the evaluated models. While it performs moderately in isolated cases like \textit{SpyBulb}, it shows very poor generalization overall. Particularly, its testing accuracy plummets for \textit{DropCam} (40.03\%) and \textit{Canary} (47.19\%), indicating its limitations in handling complex, unseen IoT camera behaviors. In summary, \textbf{DeepSVDD} stands out as the most promising solution for zero-day IoT camera detection due to its consistency and strong outlier detection, while \textbf{One-ClassSVM} appears least effective in this application domain.

\begin{table*}[hbtp]
\centering
\caption{Summary for detecting zero-day IoT (Spy) cameras (Note: x IoT Cam: y IoT Cams represents x IoT Cam used for
training model and remaining y IoT Cams used for testing models)}
\label{tab:summary_OneClassSVMipcam}
\resizebox{\linewidth}{!}{
\begin{tabular}{ccccccccccccccccc}
\hline
Models & \multicolumn{4}{c}{OneClassSVM} & \multicolumn{4}{c}{SGDOneClassSVM} & \multicolumn{4}{c}{Isolation Forest} & \multicolumn{4}{c}{DeepSVDD} \\ \hline
\multirow{2}{*}{\begin{tabular}[c]{@{}c@{}}TrainedSamples \\ : TestSamples\end{tabular}} & \multicolumn{2}{c}{Accuracy(\%)} & \multicolumn{2}{c}{Detection in testing} & \multicolumn{2}{c}{Accuracy(\%)} & \multicolumn{2}{c}{Detection in testing} & \multicolumn{2}{c}{Accuracy(\%)} & \multicolumn{2}{c}{Detection in testing} & \multicolumn{2}{c}{Accuracy(\%)} & \multicolumn{2}{c}{Detection in testing} \\ \cline{2-17} 
 & Training & Testing & \#Inliers & \#Outliers & Training & Testing & \#Inliers & \#Outliers & Training & Testing & \#Inliers & \#Outliers & Training & Testing & \#Inliers & \#Outliers \\ \hline \hline
\begin{tabular}[c]{@{}c@{}}SpyClock\\  : 10 IoT Cams.\end{tabular} & 75 & 82.70 & 20678 & 98859 & 47.50 & 64.74 & 42144 & 77393 & 90 & 98.39 & 1928 & 117609 & 97.46 & 99.06 & 1118 & 118419 \\ \hline
\begin{tabular}[c]{@{}c@{}}Canary \\ : 10 IoT Cams.\end{tabular} & 88.37 & 47.19 & 63117 & 56394 & 34.88 & 64.8 & 42073 & 77438 & 88.38 & 86.63 & 15980 & 103531 & 97.35 & 77.22 & 27227 & 92284 \\ \hline
\begin{tabular}[c]{@{}c@{}}D3D \\ : 10 IoT Cams.\end{tabular} & 90.23 & 66.26 & 39884 & 78315 & 44.25 & 65.02 & 41342 & 76857 & 89.66 & 53.73 & 54696 & 63503 & 93.52 & 52.78 & 55810 & 62389 \\ \hline
\begin{tabular}[c]{@{}c@{}}Ezviz \\ : 10 IoT Cams.\end{tabular} & 88.80 & 41.52 & 68626 & 48720 & 45.17 & 65.14 & 40912 & 76434 & 89.96 & 53.60 & 54696 & 63503 & 92.91 & 94.82 & 6084 & 111262 \\ \hline
\begin{tabular}[c]{@{}c@{}}Netatmo \\ : 10 IoT Cams.\end{tabular} & 70.53 & 54.06 & 54662 & 64327 & 1.05 & 64.50 & 42242 & 76747 & 89.47 & 28.53 & 83856 & 33490 & 96.23 & 53.16 & 55730 & 63259 \\ \hline
\begin{tabular}[c]{@{}c@{}}SpyBulb \\ : 10 IoT Cams.\end{tabular} & 89.06 & 30.26 & 80911 & 35099 & 55.73 & 70.87 & 33796 & 82214 & 89.82 & 71.89 & 33452 & 85537 & 96.32 & 93.82 & 7168 & 108842 \\ \hline
\begin{tabular}[c]{@{}c@{}}DropCam\\  : 10 IoT Cams.\end{tabular} & 88.45 & 40.03 & 67979 & 45376 & 96.50 & 95.71 & 4859 & 108496 & 89.97 & 15.99 & 97460 & 18550 & 93.34 & 78 & 25050 & 88305 \\ \hline
\begin{tabular}[c]{@{}c@{}}Netatmo\\  : 10 IoT Cams.\end{tabular} & 86.51 & 53.08 & 48208 & 54527 & 35.06 & 64.79 & 36176 & 66559 & 90.00 & 89.95 & 11505 & 101850 & 99.66 & 87.07 & 13283 & 89452 \\ \hline
\begin{tabular}[c]{@{}c@{}}SamsungSmart\\ Cam: 10 IoT Cams.\end{tabular} & 89.52 & 46.65 & 31150 & 272239 & 21.62 & 50.23 & 29063 & 29326 & 89.99 & 59.71 & 41387 & 61348 & 94.96 & 57 & 25308 & 33081 \\ \hline
\begin{tabular}[c]{@{}c@{}}TP-Link\\  : 10 IoT Cams.\end{tabular} & 87.56 & 64.79 & 38522 & 70883 & 54.70 & 66.44 & 36716 & 72689 & 90.03 & 49.19 & 29667 & 28722 & 99.98 & 56.66 & 33081 & 25308 \\ \hline
\begin{tabular}[c]{@{}c@{}}WithingsCam \\ : 10 IoT Cams.\end{tabular} & 88.08 & 70.65 & 31060 & 74784 & 38.89 & 65.35 & 36671 & 69173 & 89.99 & 83.08 & 17017 & 88827 & 94.64 & 70.06 & 73253 & 32591 \\ \hline
\end{tabular}}
\end{table*}

\subsection{Only one Zero Day}
 
\begin{table*}[hbtp]
\caption{Summary of classifiers on all IP Cameras for detecting zero-day IoT (Spy) cameras (Note: x IoT Cams: y IoT
Cam represents x IoT Cams used for training model (always 10) and y IoT Cam used for testing models (remaining 1))}
\label{tab:OneClassSVM_ipcamReverse}
\resizebox{\linewidth}{!}{
\begin{tabular}{ccccccccccccccccc}
\hline
Models & \multicolumn{4}{c}{OneClassSVM} & \multicolumn{4}{c}{SGDOneClassSVM} & \multicolumn{4}{c}{Isolation Forest} & \multicolumn{4}{c}{DeepSVDD} \\ \hline
\multirow{2}{*}{\begin{tabular}[c]{@{}c@{}}TrainedSamples\\  : TestSamples\end{tabular}} & \multicolumn{2}{c}{Accuracy(\%)} & \multicolumn{2}{c}{Detection in testing} & \multicolumn{2}{c}{Accuracy(\%)} & \multicolumn{2}{c}{Detection in testing} & \multicolumn{2}{c}{Accuracy(\%)} & \multicolumn{2}{c}{Detection in testing} & \multicolumn{2}{c}{Accuracy(\%)} & \multicolumn{2}{c}{Detection in testing} \\ \cline{2-17} 
 & Training & Testing & \#Inliers & \#Outliers & Training & Testing & \#Inliers & \#Outliers & Training & Testing & \#Inliers & \#Outliers & Training & Testing & \#Inliers & \#Outliers \\ \hline \hline
\begin{tabular}[c]{@{}c@{}}10 IoT Cams.\\  : SpyClock\end{tabular} & 90.49 & 10.38 & 354 & 41 & 35.62 & 59.24 & 161 & 234 & 90.15 & 10.63 & 353 & 42 & 95.23 & 84.81 & 60 & 335 \\ \hline
\begin{tabular}[c]{@{}c@{}}10 IoT Cams. \\ : Canary\end{tabular} & 90.32 & 42.99 & 240 & 181 & 37.09 & 37.05 & 265 & 156 & 89.99 & 57.48 & 242 & 179 & 89.89 & 43.47 & 238 & 183 \\ \hline
\begin{tabular}[c]{@{}c@{}}10 IoT Cams. \\ : D3D\end{tabular} & 89.11 & 11.89 & 1527 & 206 & 45.89 & 100 & 0 & 1733 & 90.03 & 11.94 & 1526 & 207 & 90.52 & 87.01 & 225 & 1508 \\ \hline
\begin{tabular}[c]{@{}c@{}}10 IoT Cams. \\ : Ezviz\end{tabular} & 89.76 & 33.29 & 1725 & 861 & 35.64 & 48.53 & 1331 & 1255 & 90.03 & 62.68 & 965 & 1621 & 95.47 & 76.72 & 602 & 1984 \\ \hline
\begin{tabular}[c]{@{}c@{}}10 IoT Cams. \\ : Netatmo\end{tabular} & 89.86 & 51.01 & 462 & 481 & 36.22 & 99.26 & 7 & 936 & 90.00 & 51.64 & 456 & 487 & 95.16 & 87.60 & 117 & 826 \\ \hline
\begin{tabular}[c]{@{}c@{}}10 IoT Cams. \\ : SpyBulb\end{tabular} & 88.98 & 20.32 & 3125 & 797 & 33.77 & 39.7 & 2365 & 1557 & 90.00 & 20.02 & 3137 & 785 & 99.97 & 89.39 & 416 & 3506 \\ \hline
\begin{tabular}[c]{@{}c@{}}10 IoT Cams. \\ : DropCam\end{tabular} & 91.11 & 98.02 & 132 & 6445 & 33.74 & 1.75 & 6462 & 115 & 90.00 & 1.81 & 6458 & 119 & 89.87 & 81.22 & 1235 & 5342 \\ \hline
\begin{tabular}[c]{@{}c@{}}10 IoT Cams.\\  : Netatmo\end{tabular} & 89.51 & 10.24 & 15436 & 1761 & 36.15 & 64.37 & 6127 & 11070 & 89.99 & 8.72 & 15697 & 1500 & 90.01 & 60.80 & 6740 & 10457 \\ \hline
\begin{tabular}[c]{@{}c@{}}10 IoT Cams.\\  : SamsungSmartCam\end{tabular} & 89.57 & 25.72 & 45714 & 15829 & 67.89 & 100 & 0 & 61543 & 90.00 & 23.37 & 47155 & 14388 & 90.18 & 71.61 & 17466 & 44077 \\ \hline
\begin{tabular}[c]{@{}c@{}}10 IoT Cams. \\ : TP-Link\end{tabular} & 90.51 & 0.5 & 10481 & 46 & 33.49 & 45.94 & 5691 & 4836 & 90.00 & 0.006 & 10466 & 61 & 90.05 & 60.89 & 4117 & 6410 \\ \hline
\begin{tabular}[c]{@{}c@{}}10 IoT Cams. \\ : WithingsCam\end{tabular} & 88 & 49.94 & 7052 & 7036 & 34.35 & 60.29 & 5594 & 8494 & 90.01 & 10.29 & 12638 & 1450 & 90.00 & 55.41 & 6281 & 7807 \\ \hline
\end{tabular}}
\end{table*}

Table \ref{tab:OneClassSVM_ipcamReverse} shows results of detecting zero-day IoT cameras using OCSVM, SGDOCSVM, IF and DeepSVDD when 10 IoT cameras have been used in training and only one becomes zero-day.
Among the evaluated models, \textbf{DeepSVDD} demonstrates the most robust and consistent performance, achieving the highest testing accuracy in \textbf{9 out of 11} test cases. Its strength lies in its high precision for detecting zero-day IoT devices, with particularly strong results observed on \textit{SpyBulb} (89.96\%) and \textit{D3D} (87.01\%). This indicates the model’s superior ability to generalize from a diverse set of trained devices to unseen test devices. However, a slight drop in performance is noted on \textit{Canary} (43.47\%) and \textit{TP-Link} (60.80\%), suggesting sensitivity to certain device-specific characteristics.
\textbf{Isolation Forest} delivers moderate and stable results with consistent inlier-outlier separation. It performs relatively well on devices like \textit{Canary} (57.48\%) and \textit{D3D} (62.68\%), but exhibits significantly low detection accuracy on \textit{DropCam} (1.81\%), \textit{TP-Link} (8.72\%), and \textit{SamsungSmartCam} (23.37\%), highlighting its limitations in handling diverse zero-day behaviors.
\textbf{SGDOneClassSVM} shows strong performance in specific cases, achieving high test accuracy on \textit{Ezviz} (99.26\%) and \textit{TP-Link} (95.04\%), and generally outperforming the traditional One-Class SVM in most scenarios. However, its results are inconsistent, with poor generalization on devices like \textit{DropCam} (1.75\%) and \textit{D3D} (45.89\%).
In contrast, \textbf{One-Class SVM} exhibits acceptable training performance but suffers from poor generalization capabilities. It achieves over 50\% test accuracy in only two cases—\textit{DropCam} (98.02\%) and \textit{WithingsCam} (49.94\%)—and fails to scale effectively for zero-day device detection, rendering it the least reliable model in this comparative study.

\subsection{Detecting Zero-Day IoT camera using DeepSVDD}
\begin{figure}[hbtp]
\centering
\begin{subfigure}{.40\textwidth}
    \centering
    \includegraphics[width=\columnwidth]{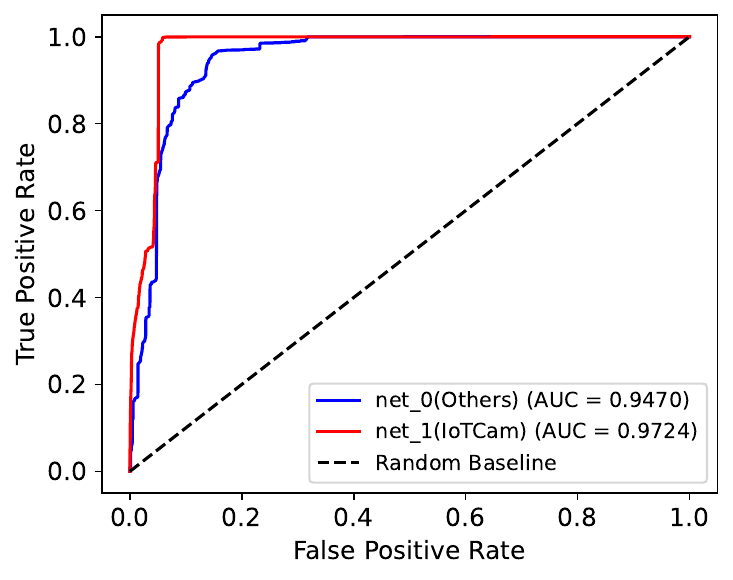}  
    \caption{RoC curve}
\end{subfigure}
\begin{subfigure}{.40\textwidth}
    \centering
    \includegraphics[width=\columnwidth]{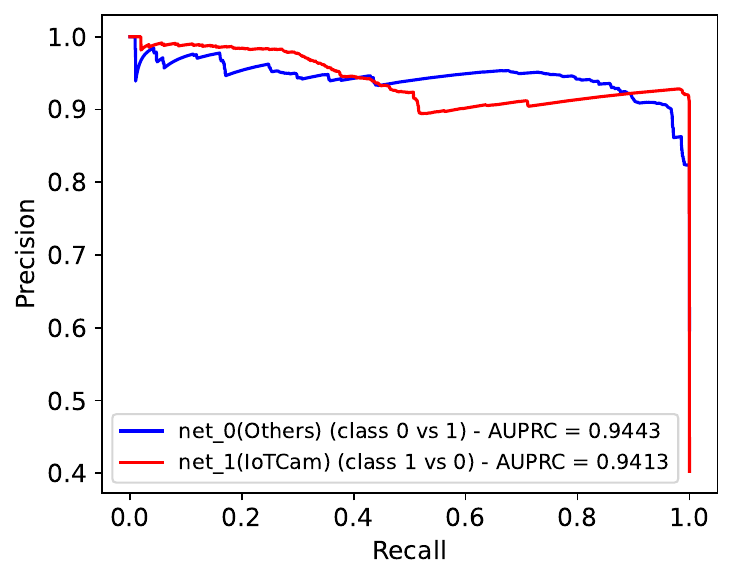}  
    \caption{Precision Recall Curve}
\end{subfigure}
\caption{Others and IoTCam datasets RoC and Precision Recall Curve using deepSVDD}
\label{fig:dsvdd_results}
\end{figure} 

We have considered a relatively advanced model of DeepSVDD for detection of zero-day IoT cameras. 
Figure \ref{fig:dsvdd_results}shows the performance of DeepSVDD using RoC Curve and precision-recall (PR) curve on Others and IoTCam datasets respectively. 
DeepSVDD learns a mapping by using a deep neural network from input space to a latent space. 
The ROC curve in Figure \ref{fig:dsvdd_results} (a) demonstrates that DeepSVDD is highly effective in anomaly detection for IoT camera data. 
With AUC values of 0.9470 (Others) and 0.9724 (IoTCam), the model is able to distinguish anomalous traffic pattern arising due to IoT cameras with high confidence. 
PR curve in Figure \ref{fig:dsvdd_results} (b) demonstrates that DeepSVDD model achieves a high precision and recall across the range of decision thresholds. 
With AUPRC values of 0.9443 (Others) and 0.9413 (IoTCam), the models turns out to be highly reliable for the identification of true anomalies while minimizing false alarms.

\section{Detecting IoT Camera Flows}
\label{sec:detection-iot-camera}

\subsection{Classifying IoT Camera Flows}
Assuming the knowledge that there may be some IoT cameras in the network, a defender may wish to classify the network flows to identify a particular IoT camera.
In this set of experiments, we consider Set I (BITSPHC) and Set II (UNSW) separately at first and then combine them together to form another dataset, let us call it as Combined. 
Recall that we have six-class and five-class classifications in Set I and Set II, respectively; consequently, eleven-class classifications in the Combined dataset. 
The results of classifications are shown in Table \ref{tab:accuracy_diffML_allfeatures}. 
XGB classifier produce an accuracy of more than 99\% in Set I; the GNB classifier shows a very low accuracy in this dataset.
Comparing the results in Set II with those in Set I, it turns out that the accuracy in Set II is slightly higher in all models. 
For instance, DT shows 99.49\% accuracy, which is about 2\% more than that in Set I. 
Among others, XGB shows consistently higher accuracy in both datasets. 
Because the datasets are collected in different scenarios, we combine the datasets to extend our experiments to a wider set of IoT cameras. 
Except GNB, the models perform better in general in the combined dataset, with XGB showing the highest accuracy of about 99.8\%.

\begin{table}[hbtp]
\centering
\caption{Accuracy of seven supervised ML Models on Set I, Set II and a Combination of Set I and Set II.}
\label{tab:accuracy_diffML_allfeatures}
\begin{tabular}{ccccc}
\hline
\multirow{2}{*}{Sr. No.} & \multirow{2}{*}{ML Models} & \multicolumn{3}{c}{\begin{tabular}[c]{@{}c@{}}Maximum Accuracy for all Features\end{tabular}} \\ 
\cmidrule(l){3-5}
 &  & BITSPHC & UNSW & Combine \\ \hline \hline
1 & DT & 97.29\% & 99.49\% & 99.56\% \\ \hline
2 & KNN & 95.68\% & 99.52\% & 99.6\% \\ \hline
3 & RF & 98.12\% & 99.71\% & 99.7\% \\ \hline
4 & GNB & 39.76\% & 65.27\% & 53.87\% \\ \hline
5 & XGB & 99.24\% & 99.84\% & 99.8\% \\ \hline
6 & ExtraTrees & 97.97\% & 99.78\% & 99.7\% \\ \hline
7 & LKSVM & 89.03\% & 97.27\% & 89.82\% \\ \hline
\end{tabular}
\end{table}

Because a defender may be further interested in the misclassifications of IoT cameras, we investigate the false positives and false negatives in this classification. 
Table \ref{tab:misclass_table} shows the results of misclassifications while classifying the IoT cameras only in Set I. The title of a column indicates the true label a flow and the cells in that column show misclassification of that camera as other cameras.
For example, 0.11\% and 0.78\% of Netatmo, i.e., the title of the first column, camera flows are misclassified as the flows of SpyBulb while using DT and XGB classifiers, respectively. 
In general, the misclassification rate is less than 1\%, the only exception is GNB. 
It turns out that the flows of Ezviz and SpyBulb get misclassified as any other IoT camera using any classifier. 
Focusing on XGB, the variations in the misclassification are relatively low. For example, 1.8\% Canary, 0.78\% Netatmo, and 0.22\% D3D flows are misclassified only as SpyBulbs, whereas 0.1\% SpyBulb flows are misclassified as Ezviz. 
Only in the case of classifying Ezviz flow variation is slightly higher, i.e., a total of less than 4\% Ezviz flows are misclassified as either Netatmo (1.8\%) or Canary (1.8\%) and SpyBulb (0.9\%). 
The result of misclassification is similar in the case of ET. 
Thus, we consider both XGB and ET to be efficient classifiers that reduce misclassification significantly. 

We show the ROC curve visualizing the performance of our classifier by plotting the True Positive Rate (TPR) against the False Positive Rate (FPR) at various threshold levels. The X-axis shows that FPR ranges from 0 to approximately 0.005. The FPR values indicate the proportion of negative instances incorrectly classified as positive. Lower values indicate a better-performing classifier. The Y-axis shows that TPR ranges from 0.943 to 0.996. The TPR values reflect the proportion of actual positive instances correctly identified by the classifier. The ROC curve shown in Fig. \ref{fig:roc_curve} rises sharply, indicating that the classifier maintains a high TPR while keeping the FPR relatively low. As FPR values close to 0 (e.g., 0.001 to 0.005), the TPR remains high, indicating that our classifier correctly identifies a large percentage of positive cases without falsely classifying many negatives. TPR values show the classifier is highly effective in detecting positive instances of IP cameras.

\begin{table*}[hbtp]
\centering
\caption{Miss-classification rate for IoT camera using different supervised ML models using Set I.}
\label{tab:misclass_table}
\resizebox{\linewidth}{!}{
\begin{tabular}{ccccccc}
\hline
Models & Netatmo & SpyClock & Canary & D3D & Ezviz & SpyBulb \\ \hline \hline
DT & SpyBulb(0.11\%) & SpyBulb(0.70\%) & \begin{tabular}[c]{@{}c@{}}SpyClock (0.30\%)\\ Ezviz (0.92\%)\\ D3D (6.19\%)\\ SpyBulb (1.23\%)\end{tabular} & \begin{tabular}[c]{@{}c@{}}SpyClock (0.17\%)\\ Netatmo (0.35\%)\\ Canary (0.17\%)\\ SpyBulb (0.53\%)\end{tabular} & \begin{tabular}[c]{@{}c@{}}D3D (1.43\%)\\ Netatmo(0.71\%)\\ Canary (4.31\%)\\ SpyBulb (3.59\%)\end{tabular} & \begin{tabular}[c]{@{}c@{}}SpyClock (0.07\%)\\ Ezviz (0.38\%)\\ D3D(0.77\%)\\ Netatmo(0.15\%)\\ Canary (0.38\%)\end{tabular} \\ \hline
GNB & \begin{tabular}[c]{@{}c@{}}SpyClock(1.89\%)\\ Ezviz(0.94\%)\\ D3D(14.51\%)\\ Canary(49.68\%)\\ SpyBulb(0.31\%)\end{tabular} & \begin{tabular}[c]{@{}c@{}}Ezviz(1.06\%)\\ D3D(5.31\%)\\ Netatmo(1.06\%)\\ Canary(62.76\%)\\ SpyBulb(6.38\%)\end{tabular} & D3D(0.93\%) & \begin{tabular}[c]{@{}c@{}}SpyClock(8.37\%)\\ Netatmo(0.45\%)\\ Canary(32.57\%)\\ SpyBulb(0.22\%)\end{tabular} & \begin{tabular}[c]{@{}c@{}}D3D(38.18\%)\\ Canary(18.18\%)\end{tabular} & \begin{tabular}[c]{@{}c@{}}SpyClock(0.19\%)\\ Ezviz(38.27\%)\\ D3D(12.82\%)\\ Netatmo(1.19\%)\\ Canary(26.73\%)\end{tabular} \\ \hline
RF & \begin{tabular}[c]{@{}c@{}}D3D(0.23\%)\\ Canary(0.11\%)\\ SpyBulb(0.47\%)\end{tabular} & \begin{tabular}[c]{@{}c@{}}Ezviz(0.72\%)\\ D3D(4.37\%)\\ SpyBulb(3.64\%)\end{tabular} & \begin{tabular}[c]{@{}c@{}}Ezviz(0.68\%)\\ D3D(0.34\%)\\ Netatmo(0.68\%)\\ SpyBulb(2.04\%)\end{tabular} & \begin{tabular}[c]{@{}c@{}}SpyClock(0.35\%)\\ Ezviz(0.53\%)\\ Canary(0.35\%)\\ SpyBulb(0.53\%)\end{tabular} & \begin{tabular}[c]{@{}c@{}}SpyClock(1.49\%)\\ Netatmo(0.74\%)\\ Canary(2.24\%)\\ SpyBulb(4.47\%)\end{tabular} & \begin{tabular}[c]{@{}c@{}}SpyClock(0.07\%)\\ Ezviz(0.15\%)\\ Netatmo(0.07\%)\\ Canary(0.22\%)\end{tabular} \\ \hline
KNN & \begin{tabular}[c]{@{}c@{}}SpyClock(0.15\%)\\ Ezviz(0.3\%)\\ D3D(0.3\%)\\ Canary(1.21\%)\\ SpyBulb(0.6\%)\end{tabular} & \begin{tabular}[c]{@{}c@{}}D3D(8.57\%)\\ Netatmo(2.85\%)\\ SpyBulb(1.9\%)\end{tabular} & \begin{tabular}[c]{@{}c@{}}Ezviz(0.42\%)\\ Netatmo(1.28\%)\\ SpyBulb(4.72\%)\end{tabular} & \begin{tabular}[c]{@{}c@{}}SpyClock(0.97\%)\\ Canary(0.24\%)\\ SpyBulb(1.45\%)\end{tabular} & \begin{tabular}[c]{@{}c@{}}SpyClock(1.78\%)\\ D3D(0.89\%)\\ Netatmo(0.89\%)\\ Canary(1.78\%)\\ SpyBulb(8.03\%)\end{tabular} & \begin{tabular}[c]{@{}c@{}}SpyClock(0.3\%)\\ D3D(0.3\%)\\ Netatmo(1.12\%)\\ Canary(2.14\%)\end{tabular} \\ \hline
XGB & SpyBulb(0.78\%) & 0\% & SpyBulb(1.8\%) & SpyBulb(0.22\%) & \begin{tabular}[c]{@{}c@{}}Netatmo(1.85\%)\\ Canary(1.85\%)\\ SpyBulb(0.9\%)\end{tabular} & Ezviz(0.1\%) \\ \hline
ET & \begin{tabular}[c]{@{}c@{}}D3D(0.46\%)\\ SpyBulb(1.24\%)\end{tabular} & \begin{tabular}[c]{@{}c@{}}D3D(5.61\%)\\ SpyBulb(0.93\%)\end{tabular} & \begin{tabular}[c]{@{}c@{}}SpyClock(0.43\%)\\ Ezviz(0.87\%)\\ D3D(0.87\%)\\ Netatmo(0.43\%)\\ SpyBulb(1.7\%)\end{tabular} & \begin{tabular}[c]{@{}c@{}}SpyClock(0.2\%)\\ Canary(0.6\%)\\ SpyBulb(0.2\%)\end{tabular} & \begin{tabular}[c]{@{}c@{}}D3D(1.76\%)\\ Canary(1.7\%)\\ SpyBulb(2.65\%)\end{tabular} & \begin{tabular}[c]{@{}c@{}}SpyClock(0.1\%)\\ Ezviz(0.1\%)\\ Netatmo(0.3\%)\\ Canary(0.3\%)\end{tabular} \\ \hline
LKSVM & \begin{tabular}[c]{@{}c@{}}SpyClock(3.55\%)\\ Ezviz(0.16\%)\\ SpyBulb(6.42\%)\end{tabular} & \begin{tabular}[c]{@{}c@{}}D3D(4.46\%)\\ SpyBulb(30.35\%)\end{tabular} & \begin{tabular}[c]{@{}c@{}}SpyClock(0.88\%)\\ Ezviz(0.88\%)\\ SpyBulb(21.33\%)\end{tabular} & \begin{tabular}[c]{@{}c@{}}Ezviz(0.42\%)\\ SpyBulb(6.42\%)\end{tabular} & \begin{tabular}[c]{@{}c@{}}SpyClock(4.58\%)\\ Canary(2.75\%)\\ SpyBulb(10.09\%)\end{tabular} & \begin{tabular}[c]{@{}c@{}}SpyClock(1.34\%)\\ Ezviz(0.2\%)\\ D3D(1.65\%)\\ Netatmo(0.82\%)\\ Canary(4.02\%)\end{tabular} \\ \hline
\end{tabular}}
\end{table*}

\subsection{Analysis of Important Features}
Analyzing the importance of features using the ExtraTree (ET) classifier, the top 10 important features have significant overlap in Set I, Set II, and the combined dataset. For example, the features of  \texttt{Init Bwd Win Byts}, \texttt{ACK Flag Cnt}, \texttt{Flow Duration}, and \texttt{Bwd IAT Tot} are common in the top 10 features (Table \ref{tab:features-select-IoT}) all these three data sets. 
Only a limited number of features are exclusively present in the top features in each of these datasets.
To check whether the top 10 features can produce comparable performance in the classification, the accuracy is slightly reduced in a model in general (results are shown in Table \ref{tab:accuracy_diffML}). 
Surprisingly, the accuracy in XGB is relatively better in UNSW and combined datasets compared to the BITSPHC dataset.
However, all other models perform better in the UNSW dataset. 
Nevertheless, the XGB and ET still outperform the other models in general. 

\begin{table}[hbtp]
    \centering
    \caption{Top 10 Features in IoT Camera using Extra Tree.}
    \label{tab:features-select-IoT}
    \resizebox{\linewidth}{!}{
    \begin{tabular}{cccc}
\hline
Rank & UNSW & BITSPHC & Combination \\ \hline \hline
1 & Bwd IAT Tot & Init Bwd Win Byts & Init Bwd Win Byts \\
2 & Init Bwd Win Byts & ACK Flag Cnt &  Bwd Header Len\\
3 & ACK Flag Cnt & Pkt Len Max &  Flow Duration\\
4 & Fwd IAT Tot & Pkt Len Min & Bwd Pkts/s \\
5 & Flow Duration & Fwd IAT Max & Fwd IAT Tot \\
6 & Pkt Len Max & Flow Duration & ACK Flag Cnt \\
7 & Bwd Pkts/s & Bwd Header Len & Bwd IAT Tot \\
8 & Fwd IAT Mean & Bwd Pkt Len Max &  Bwd Pkt Len Max\\
9 & Flow Pkts/s & Bwd Pkt Len Min & Flow Pkts/s \\
10 & Bwd PSH Flags & Fwd IAT Tot & Fwd IAT Mean \\ \hline
\end{tabular}}
\end{table}

\subsection{Comparison with Prior Works}
\begin{table*}[hbtp]
\centering
\caption{Comparison with Prior Work.}
\label{tab:comparison_accuracy}
\resizebox{\textwidth}{!}{\begin{tabular}{clcclcc}
\hline
\multirow{2}{*}{\begin{tabular}[c]{@{}c@{}}Research\\ Paper\end{tabular}} & \multicolumn{1}{c}{\multirow{2}{*}{IoT Devices}} & \multirow{2}{*}{\begin{tabular}[c]{@{}c@{}}IoT Camera \\ Detection\end{tabular}} & \multirow{2}{*}{\begin{tabular}[c]{@{}c@{}}Zero-day Spy\\ camera Detection\end{tabular}} & \multicolumn{1}{c}{\multirow{2}{*}{Techniques}} & \multirow{2}{*}{\begin{tabular}[c]{@{}c@{}}Precision\\ (\%)\end{tabular}} & \multirow{2}{*}{\begin{tabular}[c]{@{}c@{}}Recall\\ (\%)\end{tabular}} \\
 & \multicolumn{1}{c}{} &  &  & \multicolumn{1}{c}{} &  &  \\ \hline \hline
\cite{Arunan2018IEEE} & \begin{tabular}[c]{@{}l@{}}\textbf{IP Camera} (August Doorbell\\ Cam, Belkin, Canary, Dropcam,\\ Netatmo, Samsung Smart Cam,\\ TP-Link), Smart Plugs, \\ Motion Sensors, Appliances, \\ Health Monitors\end{tabular} & \xmark & \xmark & \begin{tabular}[c]{@{}l@{}}Multistage \\ Classifier\\ (Naive Bayes \\ Multinominal\\ and Random \\ Forest)\end{tabular} & -- & -- \\ \hline 
SCamF \cite{heoACSAC2022} & \begin{tabular}[c]{@{}l@{}}\textbf{IP Camera} (Yi, Xiaomi, 360, \\ TP-Link, Amcrest, goospy, \\ ieGeek, Egloo, Hej,Green,\\ JWC, Wisenet, Relohas, \\ luhoe, YINEW, Geagle)\end{tabular} & \cmark & \xmark & \begin{tabular}[c]{@{}l@{}}Network Packet \\ level information \\ (4 Features : \\ traffic volume,\\ inter packet time\\ interval, FU rate,\\ frame per\\ second (FPS))\end{tabular} & 93.5 & 1 \\ \hline
DeWiCam \cite{Cheng2018ASIACCS} & \begin{tabular}[c]{@{}l@{}}\textbf{IP Camera} (Dahua, Hikvision, \\ Yi, Xiaomi, 360,TP-Link, Nest, \\ Amcrest, D-Link, Lenovo, Haier,\\  Yoosee and non-camera traffic\end{tabular} & \cmark & \xmark & \begin{tabular}[c]{@{}l@{}}Ensemble Extra \\ Tree Classifier\end{tabular} & -- & -- \\ \hline
\textbf{zCamInspector} & \begin{tabular}[c]{@{}l@{}}\textbf{IP Camera} (Ezviz, Netatmo, \\ Canary, D3D, V380 Spy Bulb\\ Alarm Spy Clock, Drop cam,\\ TP-Link, Withings Cam, \\ Samsung Smart Cam)\end{tabular} & \textbf{\cmark} & \textbf{\cmark} & \begin{tabular}[c]{@{}l@{}}Extra Tree \\ Classifier,\\ OneClassSVM, \\ SGDOneClassSVM, \\ IsolationForest \end{tabular} & \textbf{97.75} & \textbf{1} \\ \hline
\end{tabular}}
\end{table*}

\begin{figure}[hbtp]
\centering  
\includegraphics[width=\linewidth]{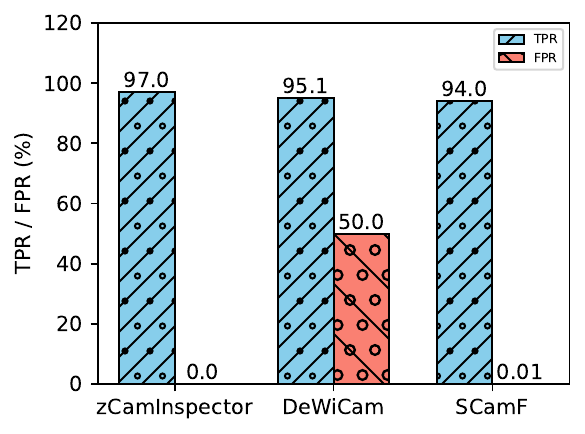}  
    \caption{Comparison of zCamInspector with Existing Work DewiCam\cite{Cheng2018ASIACCS} and SCamF\cite{heoACSAC2022} w.r.t. TPR and FPR.}
    \label{fig:tprfpr_comp}
\end{figure}

To compare the performance of existing works with our proposed zCamInspector, we look for works that are most relevant. 
Though there exists a number of works for classifying IoT devices and traditional online audio/video applications, zCamInspector is the first work that aims to classify IoT cameras only and detect zero-day IoT cameras. 
Among others, we found four existing works (\cite{Arunan2018IEEE, heoACSAC2022, singh2021USENIX, Cheng2018ASIACCS}) are relatively more relevant to our work in this paper. 
A summary of comparisons is shown in Table \ref{tab:comparison_accuracy}. 
For example, the work in \cite{Arunan2018IEEE} achieves an accuracy of more than 99\% in classifying IoT devices, including IoT cameras, smart plugs, motion sensors, and health monitors, using their own dataset.  
The work in \cite{heoACSAC2022} could achieve a precision of 0.935 with recall as 1.0 to classify IoT devices that include 20 IoT cameras (Yi, Xiaomi, 360, TP-Link, Amcrest, Goospy, ieGeek, Egloo, Hej, Green, JWC, Wisenet, Relohas, luhoe, YINEW, and Geagle) and non-camera traffic collected from Video on Demand (VOD), download, picture, game, using their own dataset. None of the existing works focuses on the detection of zero-day IoT cameras. 
The main focus of zCamInspector is to classify the IoT cameras and detect zero-day IoT cameras. 
Our proposed system could achieve the least FPR, i.e. 0, in detecting the IoT cameras. Further, IoT camera flows can be classified from non-IoT camera flows with a precision of 97.13\% and F1-Score of 97.14\%, along with an accuracy of 97.15\%. Further, to compare the performance in the same dataset as the one used in our work, we have considered only the ML model used in the four existing works. We have obtained TPR and FPR for detecting zero-day spy cameras
with 97.00\% TPR and zero FPR  (shown in Fig. \ref{fig:tprfpr_comp}).

\subsection{Comparing Runtime Performance of zCamInspector}
Since the time required to extract the features primarily depends on the duration of a flow, we look into the distribution of flow duration across the applications. 
The distribution of \texttt{flow duration} varies between the range [0, 119]sec, having a mean of 25.23s with a standard deviation of 40sec.
The time consumed for extracting the features from a flow of any duration is roughly constant, and this is mainly because all the modules in \texttt{CICFlowmeter} need to be executed whenever the flow has two or more packets. 
Investigating the runtimes in the model training and testing shown in Table \ref{tab:run_per_all_testing}). 
\texttt{zCamClassifier} consumes slightly higher time than \texttt{zCamDetector}, and this is about 2.4 sec in training and 0.19 sec in testing with small standard deviations. Overall, both the training and testing processes are significantly faster than the average duration of flows in any application. 

\begin{table}[hbtp]
    \centering
\captionof{table}{Comparison of runtime of zCamInspector with existing works during training and testing for supervised classifications.}
\label{tab:run_per_all_testing}
\begin{tabular}{ccccc}
\hline
\multirow{2}{*}{Application} & \multicolumn{2}{c}{\begin{tabular}[c]{@{}c@{}}Mean (Sec)\end{tabular}} & \multicolumn{2}{c}{\begin{tabular}[c]{@{}c@{}}Std Dev (Sec)\end{tabular}} \\ \cmidrule(l){2-5} 
 & Train & Test & Train & Test \\ \hline \hline
ZCamDetector & 1.06 & 0.29 & 0.03 & 0.01 \\ \hline
zCamClassifier & 2.4 & 0.19 & 0.1 & 0.1 \\ \hline
\end{tabular}
\end{table}


\begin{table}[hbtp]
    \centering
 \caption{Accuracy of different ML Models on datasets using top 10 features}
    \label{tab:accuracy_diffML}   
\resizebox{\linewidth}{!}{
\begin{tabular}{ccccc}
\hline
\multirow{2}{*}{Sr. No.} & \multirow{2}{*}{ML Models} & \multicolumn{3}{c}{\begin{tabular}[c]{@{}c@{}}Maximum Accuracy for Top 10 Features\end{tabular}} \\ 
\cmidrule(l){3-5}
 &  & BITSPHC & UNSW & Combine \\ \hline \hline
1 & DT & 92.82\% & 97.41\% & 96.28\% \\ \hline
2 & KNN & 92.04\% & 97.25\% & 95.89\% \\ \hline
3 & RF & 92.91\% & 96.97\% & 95.68\% \\ \hline
4 & GNB & 24.04\% & 40.57\% & 36.58\% \\ \hline
5 & XGB & 93.72\% & 97.47\% & 96.46\% \\ \hline
6 & ExtraTrees & 92.75\% & 97\% & 95.5\% \\ \hline
7 & LKSVM & 70.63\% & 85.87\% & 78.95\% \\ \hline
\end{tabular}}
\end{table}

\subsection{Ablation Study of zCamInspector}
\begin{figure}[hbtp]
\centering  
\includegraphics[width=\linewidth]{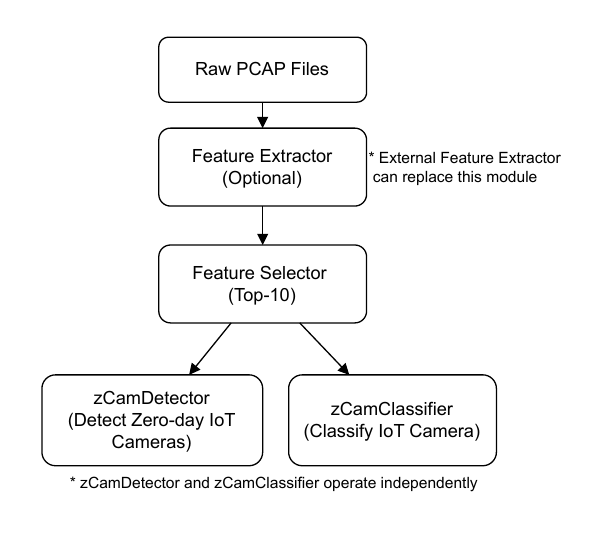}  
    \caption{Modules of zCamInspector}
    \label{fig:modulezCamInspector}
\end{figure}

zCamInspector has four major modules, namely feature extractor, feature selector, zCamDetector and zCamClassifier (shown in Fig. \ref{fig:modulezCamInspector}). 
The feature extractor module is responsible for extracting flow-based features from the raw network packets. 
This module is essential if a defender is not using any other machine learning based solution, in other words, the pcap files are not processed in any other ways. 
Alternately, if the defender has already an external module that can extract the features required for our proposed system, this module can be bypassed and the feature set can be directly taken as input to the feature selector module. 

The feature selector module is essential in zCamInspector due to the varying requirements of the models. 
Our analysis shows that the detection of zero-day IoT cameras can be done based on most important ten features. Note that the clustering using either PCA or GMM has been an additional set of experiments that helped us to understand the feasibility of discovering the traffic of IoT cameras from the non-camera devices. 
Each of four one class classifiers, i.e., OCSVM, SGDOCSVM, IF and DeepSVDD, shows its effectiveness for the detection of zero-day IoT cameras. However, a defender can used either IF or DeepSVDD as these two models have relatively better performance using a reduced set of features. 

We envision that a future IoT network can include only a limited set of IoT cameras in the network, for example, to comply with organizational policy. 
Hence, zCamDetector is essential in the system. This module works independent of zCamClassifier. 
zCamClassifier is an important module that classifies one particular camera from another. This module is required when a defender wants to ensure certain access control policy for the IoT cameras. 
This module need not rely on the output of zCamDetector, but it is useful to check if a camera is not part of the network already. 

\subsection{Limitation of the Approach}
While the performance of zCamInspector is commendable, we see a number of limitations of this approach. 
First, the data set belonging to IoT devices offering spy camera service is limited, particularly in the presence of a large number of commercially available IoT cameras.
Specifically, we have used only two spy IoT cameras, namely the V380 spy bulb and alarm cum spy camera, whereas a large number of commercial IoT devices are available at low cost. 
We notice that the inclusion of network traffic of these devices can affect the performance of zCamInspector.  
Second, Identifying the video and non-video streams from an IoT Spy Camera can be challenging due to the integrated nature of its services.
These limitations may lead to more investigations of both a better model, like MLP, or a better set of features.  

\section{Conclusion}
\label{sec:conclusion}
To address the problem of identifying IoT (spy) cameras and detecting zero-day IoT cameras, we have designed and developed a system called \textit{zCamInspector}. 
A network administrator can use zCamInspector as a first-level defense, where it can identify an IoT camera by using the patterns of network flows of the IoT (spy) cameras, or it can detect zero-day IoT cameras within its network without relying on any IP addresses or transport ports. 
To investigate the efficacy of zCamInspector, we have used a total of about 40GB of network traces including both open-sourced data and the data in our own laboratory setup to demonstrate the efficacy of our system outside our own laboratory setup. zCamInspector uses seven supervised classifiers, namely DT, kNN, RF, GNB, XGB, ExtraTree, and LKSVM, for identifying the IoT (spy) camera, and our results show that XGB with more than an average of 98\% accuracy outperforms all other models in all the data sets. 
Our analysis of features using the ExtraTree classifier shows that the top 10 features in the three datasets have non-negligible differences and hence we consider our own network trace as a new IoT network trace for the community.  
Our analysis has shown that XGB achieves both a higher precision of more than 97\% and a higher TPR compared to the state-of-the-art. 
To detect zero-day IoT cameras, zCamDetector uses four one class classifiers, namely OCSVM, SGDOCSVM, IF and DeepSVDD.
Using Set III (i.e., \enquote{Others} - the dataset representing non-IoT camera devices),  we have shown that DeepSVDD outperforms among OCSVM,  IF and SGDOCSVM to detect zero-day IoT cameras when none of the IoT cameras are seen before. 
Further, we have shown that zCamDetector can effectively detect the zero-day IoT cameras even if some the IoT cameras are already part of the network.  
It achieves best performance, using OCSVM, while detecting any of the IoT cameras other than DropCam. 
In general, our results show a promising result of any zero-day IoT camera using DeepSVDD, compared to the other three one class classifiers. In future work, we plan to extend zCamInspector to adaptive learning settings, enabling it to update its model incrementally as new devices are discovered, and to evaluate its performance in large-scale smart homes or industrial IoT environments with dynamic device churn and adversarial behaviors.

\bibliographystyle{IEEEtran}
\bibliography{references}

\end{document}